\documentclass[%
 sor,
 jor,
 amsmath,amssymb,
 reprint,%
]{revtex4}

\usepackage {setspace}
\usepackage{graphicx}
\usepackage[english]{babel}
\usepackage{dcolumn}
\usepackage{bm}
\usepackage{tikz}
\usepackage{graphicx}
\usepackage{amssymb}
\usepackage{amsmath}
\usepackage{bm}
\usepackage{enumitem}
\usepackage{etoolbox}
\begin{document}

\preprint{AIP/123-QED}

\title{Effect of adhesive interaction on strain stiffening and dissipation in granular gels undergoing yielding}

\author{Sebanti Chattopadhyay$^a$}%
\author{Sharadhi Nagaraja$^{a,}$$^b$}

\author{Sayantan Majumdar$^a$}
\thanks{smajumdar@rri.res.in}
\affiliation{$^a$Soft Condensed Matter Group, Raman Research Institute, Bengaluru 560080, India}
\affiliation{$^b$Department of Physics and Materials Science, Universit$\acute{e}$ du Luxembourg, L-1511 Luxembourg}

\date{\today}

\begin{abstract}
Stress induced yielding/fluidization in disordered solids, characterized by irreversibility and enhanced dissipation, is important for a wide range of industrial and geological processes. Although, such phenomena in thermal systems have been extensively studied, they remain poorly understood for granular solids. Here, using oscillatory shear rheology and in-situ optical imaging, we study energy dissipation in a dense granular suspension of adhesive particles that forms yield stress solids far below the isotropic jamming point obtained in the limit of hard-sphere repulsion. We find interesting non-linear flow regimes including intra-cycle strain stiffening and plasticity that strongly depend on the applied strain amplitude ($\gamma_0$) and particle volume fraction ($\phi$). We demonstrate that such nonlinearity over the entire parameter range can be effectively captured by a dimensionless variable termed as the normalized energy dissipation ($E_N$). Furthermore, in-situ optical imaging reveals irreversible particle rearrangements correlating with the  spatiotemporal  fluctuations in local velocity, the nature of which strikingly varies across the yielding transition. By directly measuring the critical jamming packing fractions using particle settling experiments, we propose a detailed phase diagram that unravels the role of inter-particle interactions in controlling the flow properties of the system for a wide range of $\gamma_0$ and $\phi$ values.
\end{abstract}

\maketitle


\section{Introduction}
Diverse disordered materials close to jamming show a finite elasticity under small perturbations. When the applied perturbations become large enough, plasticity and irreversible deformations take place as the material yields \cite{Falk2011Feb, Berthier2011Jul, Bonn2017Aug, Coussot2014Sep}. 
Yielding in amorphous solids plays an important role in material processing industries as well as catastrophic natural phenomena like landslides and earthquakes. Yielding also implies an enhanced energy dissipation \cite{vanderVaart2013Jun, Bonn2017Aug} that eventually leads to fluidization/fracture of the material. As opposed to crystalline systems, due to lack of translational symmetry there is no obvious structural precursor (similar to crystalline defects) of yielding for  disordered materials \cite{patinet2016connecting, Richard2020Nov}. This makes the understanding of yielding phenomena in amorphous solids particularly challenging. 
\newline
\newline
In recent years there have been extensive studies, both experimental and theoretical, to understand failures in amorphous materials \cite{Lin2014Oct, Bonn2017Aug, koumakis2011two, pham2006yielding, grenard2014timescales, Leishangthem2017Mar, Karmakar2010Nov, Maloney2006Jul, Keim2013, Jaiswal2016Feb, Regev2015Nov, Knowlton2014Aug, Shrivastav2016Oct, nicolas2018deformation, sastry2020models, barlow2020ductile, donley2019time, liu2021elastoplastic}. Many of these studies are motivated by the generalized jamming phase diagram describing yielding in a variety of disordered materials based on only a few control parameters \cite{Liu1998Nov}. Dense particulate suspensions have been widely used as model systems for studying the yielding behaviour in such systems. Depending on the inter-particle interactions, these systems show striking non-linear flow properties like, yielding, shear-thinning, shear-thickening, shear induced jamming \cite{Fall2008Jan, Seto2019Aug, Vazquez-Quesada2016Aug, brown2014shear, Cheng2011Sep, Dhar2019Dec}.   
\newline
\newline
Although, yielding and failure in glassy and gel-like materials formed by Brownian particles have been widely studied, experiments probing the same in non-Brownian granular suspensions of rigid particles are very few. Besides being inherently out of equilibrium, the main complexity arises from the fact that due to the large size of the constituent particles, flow of granular materials are dominated by contact interactions, since the stress scale for contact formation between two repulsive particles, $\sigma* \propto 1/d^2$  ($d$: particle diameter) \cite{guy2015towards}. Moreover, close to jamming, the average surface to surface separation between two particles becomes negligible compared to the particle size. As a result, van der Waals and other short-ranged attractive/adhesive interactions become essentially inevitable in granular systems \cite{Koeze2018Nov}. All these contributions make the flow behaviour of these systems extremely complicated. 
\newline
\newline
To cut through the complexity arising due to the frictional and adhesive/attractive interactions between non-Brownian particles, recently proposed constraint rheology models have been quite successful \cite{guy2018constraint, Richards2020Mar, Richards2020Nov, Singh2020Jun}. In these models, all the interactions are encoded in the jamming packing fraction for the system, which gradually decreases as the sliding and rolling degrees of freedom for the particles become more and more constrained due to enhanced inter-particle interactions. Such formulation, together with a generalized version of Wyart-Cates model \cite{wyart2014discontinuous}, successfully captures many aspects of yielding in granular systems. The complex interplay between inter-particle adhesion and friction is shown in a very recent study \cite{Richards2020Mar} that not only highlights the difference in the yielding behavior of non-Brownian systems and colloidal gels, but also demarcates the microscopic nature of yielding in adhesive and attractive systems. 
\newline
\newline
Despite these limited number of recent studies, many important aspects of yielding phenomena in adhesive non-Brownian suspensions formed by rigid frictional particles still remain unexplored. Stress induced viscoelastic deformation of fractal clusters formed due to adhesive inter-particle interactions can also play a key role even when the total number of adhesive/frictional contacts in the system remains fixed. Such deformations can reflect as quasi-reversible strain stiffening response in bulk rheological measurements as has also been observed in colloidal gels \cite{colombo2014stress, vanDoorn2018May}. Importantly, in granular systems, deformations of fractal particle clusters involve both adhesive and frictional contacts between the particles. Since, the number of these contacts depends on both applied perturbation and particle volume fraction, the strain stiffening response should also depend on these parameters. However, a detailed study of strain stiffening and its correlation with inter-particle interactions in the context of yielding in adhesive granular systems is currently lacking. Understanding the origin of such non-linearities also remains beyond the existing theoretical models for these systems. 
Furthermore, athermal systems of attractive/adhesive particles show significant strain-localization which gets enhanced with increasing strength of interaction, as shown by recent experimental and simulation studies \cite{Becu2006Apr, Liberto2020Oct, Irani2014May, Irani2016Nov, Chaudhuri2012Feb, vasisht2020emergence, singh2020brittle}. Thus, a complete picture of the yielding behavior requires the bulk non-linearity and energy dissipation to be connected to the local strain distribution, particle-scale interactions and irreversibility in these systems. 
\newline
\newline    
Here, we address these issues by studying non-linear mechanics, energy dissipation and strain localization in soft solids formed by dense suspensions of corn-starch particles in paraffin oil over a wide range of particle volume fractions and applied strain amplitudes. By tuning the strength of inter-particle adhesion using surfactant, we clearly identify the origin of the observed non-linear strain stiffening and strain localization. Remarkably, we directly measure the critical jamming packing fractions using particle settling experiments that encode the inter-particle interactions in the system. We show that these critical parameters can explain the complete flow behaviour of the system as described in the detailed phase diagram.  

\section{Results and Discussion}
Dense suspensions are prepared by dispersing corn-starch (CS) particles in paraffin oil for different volume fractions $\phi$ (Materials and Methods). CS particles are rigid amorphous particles having mean-diameter $\approx$ 15 $\mu$m with a polydispersity of $\approx$ 0.3 (Fig. S1). We show the variation of Elastic ($G'$) and Viscous ($G''$) moduli as a function of strain amplitude ($\gamma_0$) for $\phi$ = 0.4 in Fig. 1a. The system remains predominantly elastic ($G' > G''$) till intermediate $\gamma_0$. However, for larger strain amplitudes, a crossover to fluidization is observed with $G''> G'$. Interestingly, we do not find a linear viscoelastic region even for the smallest value of $\gamma_0$ that we use. Such non-linear behaviour is further confirmed by significant contribution of the higher harmonics (Fig. S2). To study the effect of inter-particle interactions in controlling the mechanical properties of the system under small perturbations, we plot $G'$ values averaged over strain amplitude range $0.001 < \gamma_0 < 0.005$ (shaded region in Fig. 1a) for different $\phi$ values in case of both adhesive (CS in oil) and repulsive (CS in water) systems in Fig. 1b. We find that the adhesive interactions give rise to significant elasticity at $\phi$ values much lower compared to that required for the repulsive system to have a similar elasticity, as has been observed in numerical simulations \cite{Irani2014May} and other experimental studies \cite{Trappe2001Jun}. For adhesive inter-particle interactions, system spanning networks comprised of fractal aggregates (Fig. S3) impart stability to the system for average coordination numbers well below the Maxwell isostaticity criterion \cite{Maxwell1864Apr}. Using confocal imaging (Materials and Methods) we indeed observe such a system spanning porous structure inside a stable bed (specific gravity for CS $\approx$ 1.6 and that for paraffin oil $\approx$ 0.89) settled under gravity (Inset, Fig. 1b and also Movie 1, Movie 2). These porous structures are stabilized by the adhesive interactions to support their own weight which also explains the origin of bulk elasticity in the system for volume fractions far below the repulsive random close packing limit ($\approx$ 0.56) \cite{Peters2016Apr}. As expected, we see that such porous structures are not stable under gravity in case of repulsive interparticle interactions (Movie 1, Movie 2). 
\begin{figure}
    \begin{center}
    \includegraphics[height = 10.2 cm]{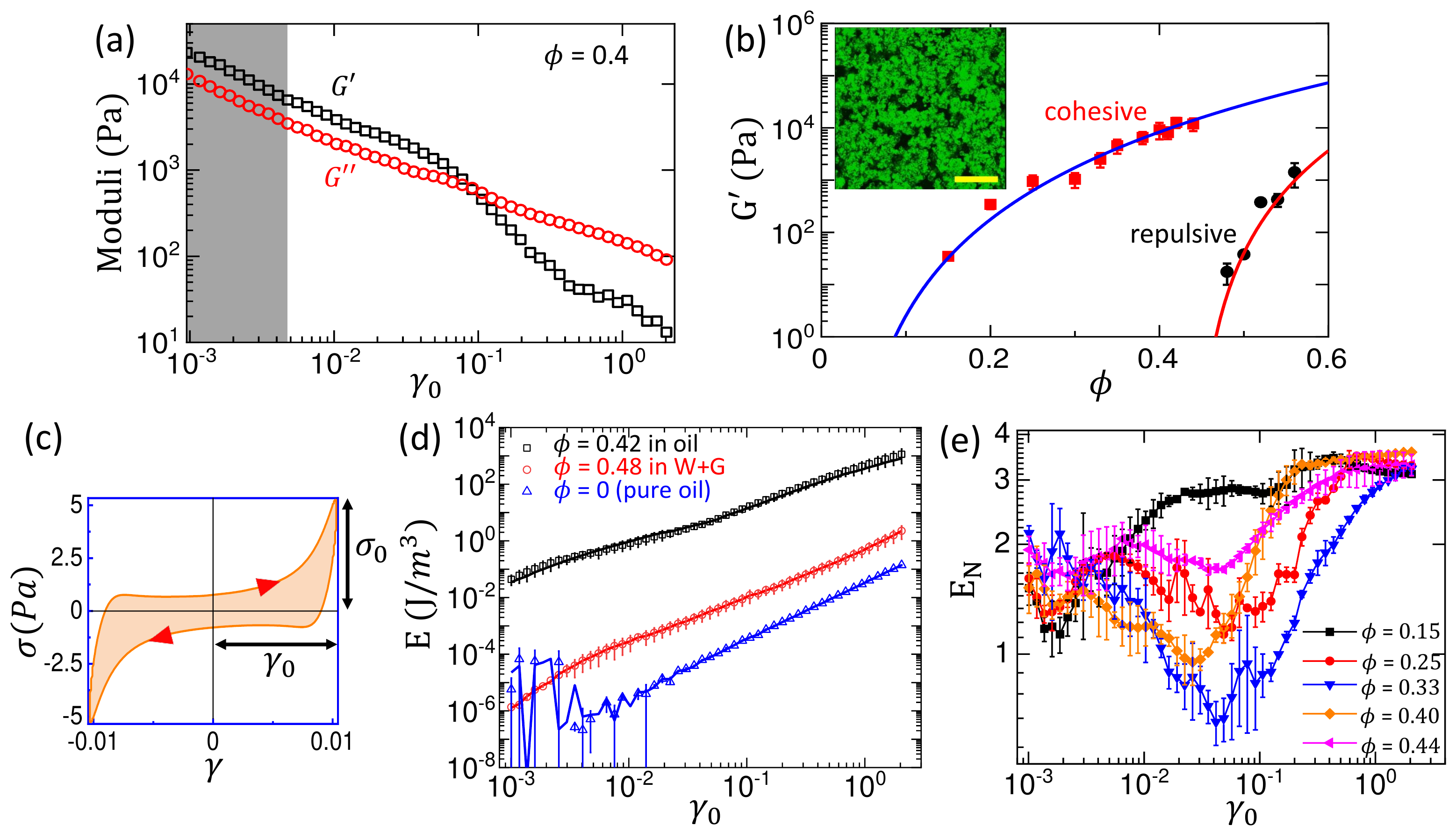}
    \caption{(a) Elastic ($G'$) and viscous ($G''$) moduli as a function of applied strain amplitude $\gamma_0$ for cornstarch particles (CS) dispersed in Paraffin oil for a volume fraction ($\phi$) $=$ 0.4. (b) $G'$ vs. $\phi$ for CS in water (black circles) and CS in Paraffin oil (red squares). The error bars represent the standard deviation of $G'$ values over the range of $\gamma_0$ indicated by the shaded region in panel (a). The solid lines are fits to the empirical relation $G' = G_{\phi}(\phi -\phi_J)^{\alpha}$. $\alpha$ = 5.1, $\phi_J$ = 0.02 for CS in oil and $\alpha$ = 4.6, $\phi_J$ = 0.44 for CS in water. Inset shows maximum intensity projection of typical confocal z-stack images for CS in oil system inside a gravitationally settled bed. Scale bar: 150 $\mu$m. (c) A typical elastic Lissajous plot (orange curve) obtained for the data shown in panel (a). The peak stress $\sigma_0$ and strain amplitude $\gamma_0$ for the Lissajous plot are also indicated. Blue bounding box denotes the Lissajous plot for an ideal plastic material for the same $\sigma_0$ and $\gamma_0$. (d) Variation of dissipated energy density ($E$) as a function of $\gamma_0$ for CS in oil ($\phi$ = 0.42), CS in water-glycerol mixture ($\phi$ = 0.48)  and pure solvent (oil) as indicated by different symbols. The solid lines indicate $E = \pi\,\gamma_0^{2}\,G''$. (e) Normalized energy dissipation ($E_N$) as a function of $\gamma_0$ for different $\phi$ values indicated in the figure. For (d) and (e) the 	error bars are the standard deviation from two independent measurements.}
    \label{F1}
    \end{center}
\end{figure}

\begin{figure}
    \begin{center}
    \includegraphics[height = 5.2 cm]{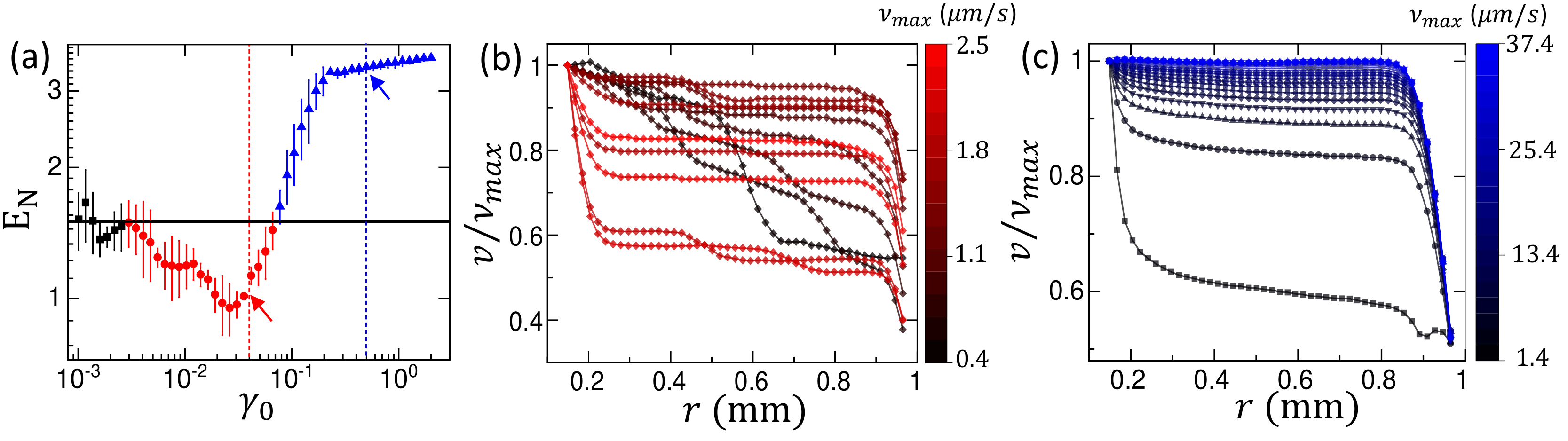}
    \caption{(a): Normalized energy dissipation $E_N$ vs strain amplitude $\gamma_0$ for $\phi$ = 0.4. The horizontal black line indicates $E_N = E_N^0$ the average small strain value of $E_N$. The points for which $E_N < E_N^0$ and $E_N > E_N^0$ are marked with red circles and blue triangles, respectively. The arrows indicate the $\gamma_0$ values for which the velocity profiles are shown in (b) and (c). The error bars are the standard deviation from two independent measurements. Velocity profiles across the gap between the two plates normalized by the instantaneous maximum velocity of the sample (near the moving plate beyond the plate roughness) for $\gamma_0$ = 0.04 (panel (b)) and $\gamma_0$ = 0.5 (panel (c)). The color gradients indicate the instantaneous maximum velocity in the sample (approximately equal to the plate velocity at that instant) during one quarter of an oscillatory applied strain cycle.
}
    \label{F2}
    \end{center}
\end{figure}
We find that the $G'$ varies as a function of $\phi$ as a power-law: $G' = G_{\phi}(\phi -\phi_J)^{\alpha}$ for both adhesive and repulsive systems (Fig. 1b) as observed earlier \cite{Trappe2001Jun}. We also record intra-cycle stress ($\sigma$) vs strain ($\gamma$) for each data point in Fig. 1a to obtain Lissajous plots for different strain amplitudes. One such plot is depicted in Fig. 1c showing a clear signature of non-linear strain stiffening when the differential shear modulus $K = \frac{d\,\sigma}{d\,\gamma}$ increases with increasing strain. The area enclosed by the Lissajous plot gives the intra-cycle energy dissipation per unit volume (dissipated energy density): $E = \oint\,\sigma(\gamma)\,d\gamma$. We compare the variation of $E$ as a function of $\gamma_0$ for both adhesive (CS in oil) and repulsive systems (CS in water-glycerol mixture) in Fig. 1d. We observe that $E$ increases monotonically with increasing $\gamma_0$ and $\phi$ values (also see Fig. S4). Importantly, we see that the dissipated energy for the adhesive system remains several orders of magnitude higher than that for the repulsive system over the entire strain range. This implies that the adhesive interactions hugely enhance the dissipation in the system over the entire range of $\gamma_0$. In all cases, we find that the analytical expression $E = \pi G''\gamma_0^2$ captures the variation of the dissipated energy accurately (Fig. 1d). However, in our case $G''$ is not a constant but decreases with $\gamma_0$ roughly as a power law over the entire range of strain amplitudes (Fig. 1a). This implies $E \sim \gamma_0^{\alpha}$ with the exponent $\alpha < 2$, unlike the case of pure viscous or linear viscoelastic materials ($\alpha = 2$) \cite{vanderVaart2013Jun}. 
\begin{figure}
    \begin{center}
    \includegraphics[height = 10.5 cm]{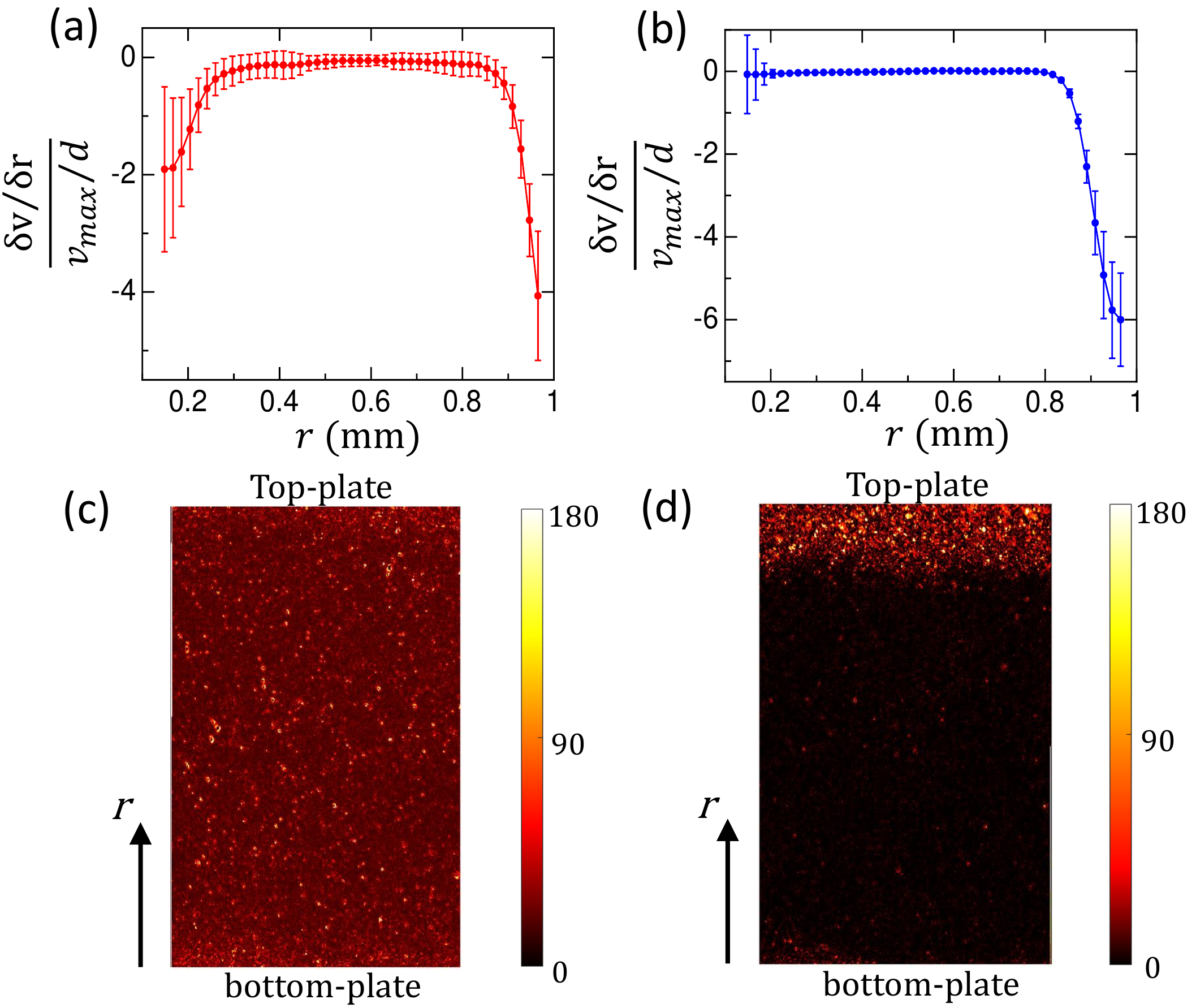}
    \caption{(a) and (b) indicates the mean dimensionless velocity gradient across the shearing gap in the system ($\phi$ = 0.4) over a complete cycle of oscillation for $\gamma_0$ =  0.04 and $\gamma_0$ =  0.5, respectively. The error bars quantify the standard deviation of local gradient fluctuations at different instances of time during the applied strain cycle. (c) and (d) indicates the stroboscopic difference between images captured at $\gamma(t)$ and $\gamma(t + T)$ ($T$: time period of the applied oscillatory strain) for $\gamma_0$ =  0.04 and $\gamma_0$ =  0.5 respectively.
}
    \label{F3}
    \end{center}
\end{figure}

\begin{figure}
    \begin{center}
    \includegraphics[height = 11 cm]{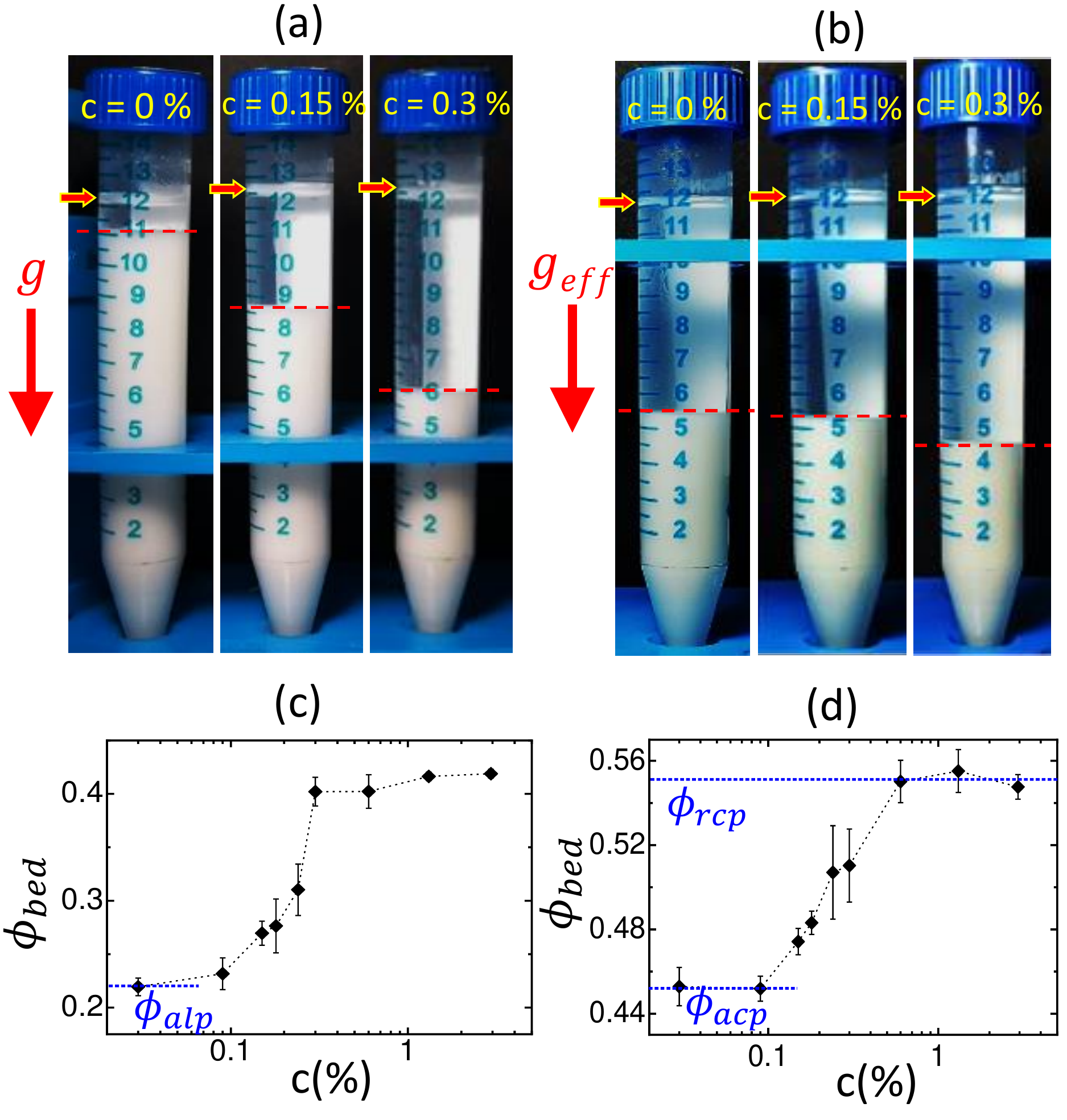}
    \caption{(a) Stable settled bed heights obtained under gravity for different surfactant concentrations ($c$) as indicated in the figure. (b) Same obtained under centrifugation at a rotation rate of 2000 rpm ($g_{eff} = 455g$, $g$: acceleration due to gravity). Variation of the packing fraction inside the settled bed ($\phi_{bed}$) under gravity (panel (c)) and under centrifugation at 2000 rpm (panel (d)) with increasing surfactant concentrations. In all cases the initial volume fraction of the sample is 0.2. The value of  $\phi_{bed}$ at zero surfactant concentration under gravitational settling gives $\phi_{alp}$ as shown in (c). $\phi_{bed}$ at zero surfactant concentration for centrifugal settling gives $\phi_{acp}$ and for high surfactant concentration at saturation gives $\phi_{rcp}$ as indicated in (d). In (c) and (d) the error bars are the standard deviation for four independent measurements. 
}
    \label{F4}
    \end{center}
\end{figure}
We find that dissipated energy $E$ does not provide much information about the non-linearity in the system as a function of $\gamma_0$ and $\phi$ values as indicated by the self-similar nature of the curves in Fig. S4. Thus, to capture the non-linearity and dissipation in the system over a wide range of parameters, we define normalized energy dissipation, $E_N = \frac{E}{\sigma_0\,\gamma_0}$, where $\sigma_0$ is the peak stress corresponding to an applied strain amplitude $\gamma_0$. Physically, the quantity $4\sigma_0\gamma_0$ denotes the dissipation for an ideal plastic material as indicated by the blue bounding box in Fig. 1c. We show the variation of $E_N$ as a function of $\gamma_0$ for a range of $\phi$ values (Fig. 1e) where we find a non-monotonic variation of $E_N$ as a function of both $\gamma_0$ as well as $\phi$. A decrease in $E_N$ with increasing $\gamma_0$ implies a strain stiffening behavior.
\\\\ 
To understand the origin of this non-monotonicity of $E_N$, we look at the flow behaviour of the system by mapping out the velocity profiles in the sample during the applied oscillatory deformations. Fig. 2a shows the typical variation of $E_N$ as a function of $\gamma_0$ for CS in oil system ($\phi = 0.4$). For small $\gamma_0$ values, the average $E_N$ is denoted by $E_N^0$. With increasing $\gamma_0$, the value of $E_N$ at first becomes smaller than $E_N^0$, then at higher $\gamma_0$ values, $E_N$ becomes larger than $E_N^0$ and finally saturates for $\gamma_0 \geq 0.2$ (Fig. 2a). In the region where $E_N < E_N^0$, the system displays strain stiffening response which disappears in the region with $E_N > E_N^0$, where the mechanical response shows strain-weakening/plasticity. The value of $\gamma_0$ beyond which $E_N$ becomes larger than $E_N^0$ marks the yield strain in our case. In Fig. 2b and 2c, we show the normalized velocity profiles across the gap when the velocity of the moving plate increases from zero to the maximum (which constitutes one quarter of a full cycle of strain deformation) for $\gamma_0$ = 0.04 and $\gamma_0$ = 0.5, respectively. In both cases we observe strong shear-banding as also observed for other adhesive/attractive systems \cite{Becu2006Apr, Liberto2020Oct, Irani2014May}. Also, for $\gamma_0$ = 0.04 we observe random spatio-temporal fluctuations in the velocity (Fig. 2b), whereas for $\gamma_0$ = 0.5 such velocity fluctuations are absent and the velocity profiles become much more self similar, particularly, for the higher plate velocities (Fig. 2c).
\\\\ 
We now focus on the spatio-temporal distribution of velocity gradients in the system over a complete cycle of strain deformation. Since we apply an oscillatory strain the boundary velocity is not fixed but continuously changing in a sinusoidal manner. For this reason we define a dimensionless gradient given by the ratio of the local velocity gradient and the average velocity gradient across the gap (assuming an affine deformation) measured at the same instant of time: $\frac{\delta \,v(r)/\delta \,r}{v_{max}/d}$. In Fig. 3a and 3b, we show for $\phi = 0.4$ the mean dimensionless velocity gradient across the shearing gap over one complete cycle of strain deformation for $\gamma_0$ = 0.04 and 0.5, respectively. The error bars denote the standard deviation of the velocity gradient fluctuations at different instants of time within a particular shear-cycle. We see from Fig. 3a and 3b that the velocity gradients in the system vary randomly across the entire gap in the strain stiffening region ($\gamma_0$ = 0.04), but such fluctuations get localized near the boundaries for larger strain values in the plastic deformation region ($\gamma_0$ = 0.5). Remarkably, both below and above yielding, the velocity gradient over a significant portion of the sample (away from the shearing boundaries) remains negligible. This indicates that across the fluidization/yielding transition, there are coexisting solid-like and fluid-like regions inside the sample. Even deep inside the fluidization regime at large $\gamma_0$ values, such coexistence remains with the fluidized regions confined near the shearing boundaries. Thus, our experiments point out spatially heterogeneous nature of the flow in adhesive granular system both above and below yielding. It is interesting to note that, the magnitude of mean velocity gradient near the top plate is stronger than that near the bottom plate (Fig. 3a and 3b) as also observed for other $\phi$ values. Although, at high volume fractions ($\phi >$ 0.22) the system forms a yield stress solid, there can still be a slight asymmetery in the strength of inter-particle contacts induced by the gravitational stress due to the higher specific gravity of CS ($\approx$ 1.6) as compared to that of paraffin oil ($\approx$ 0.89). Owing to the slightly weaker contacts, the suspension near the top plate gets fluidized relatively easily under an applied shear. Next, we look at the irreversible particle reorganizations in the system by calculating stroboscopic image difference. Basically, for a given $\gamma_0$, we calculate the difference between two gray-scale images of the sample boundaries across the shear gap captured at time $t$ and $t + T$, where $T$ is the time period of the applied oscillatory stain (Fig. S5). If the strain deformation inside the sample is completely reversible, such difference image should show zero intensity (within the dark-noise limit of the camera) everywhere. On the other hand, finite intensity in specific locations indicates irreversible particle rearrangements (localized plastic deformations) in the system. We see from Fig. 3c and 3d that such irreversible rearrangements take place uniformly throughout the sample below yielding, but such events get strongly localized near the shearing boundary for larger strain values above yielding. Remarkably, this behaviour is strongly correlated with the velocity gradient distribution in the system (Fig. 3a and 3b): the plastic rearrangements predominantly take place at spatial positions where the velocity gradients show significant spatio-temporal fluctuations.  
\begin{figure}
    \begin{center}
    \includegraphics[height = 6 cm]{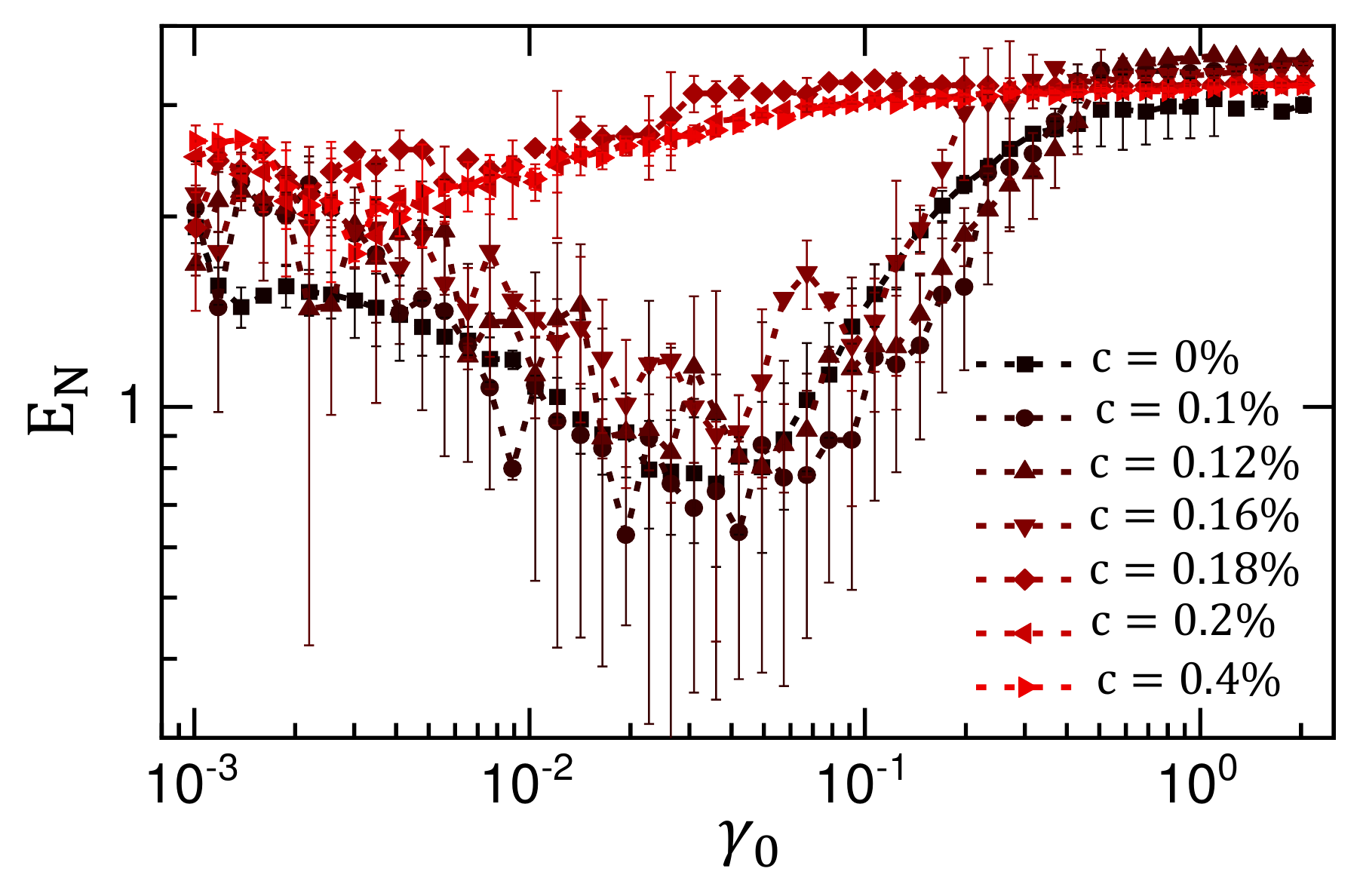}
    \caption{ Variation of normalized energy dissipation $E_N$ as a function of strain amplitude $\gamma_0$ for $\phi$ = 0.35 for increasing surfactant concentrations as indicated. Here, the error bars indicate standard deviation for two independent experimental runs. 
}
    \label{F5}
    \end{center}
\end{figure}
\begin{figure}
    \begin{center}
    \includegraphics[height = 10 cm]{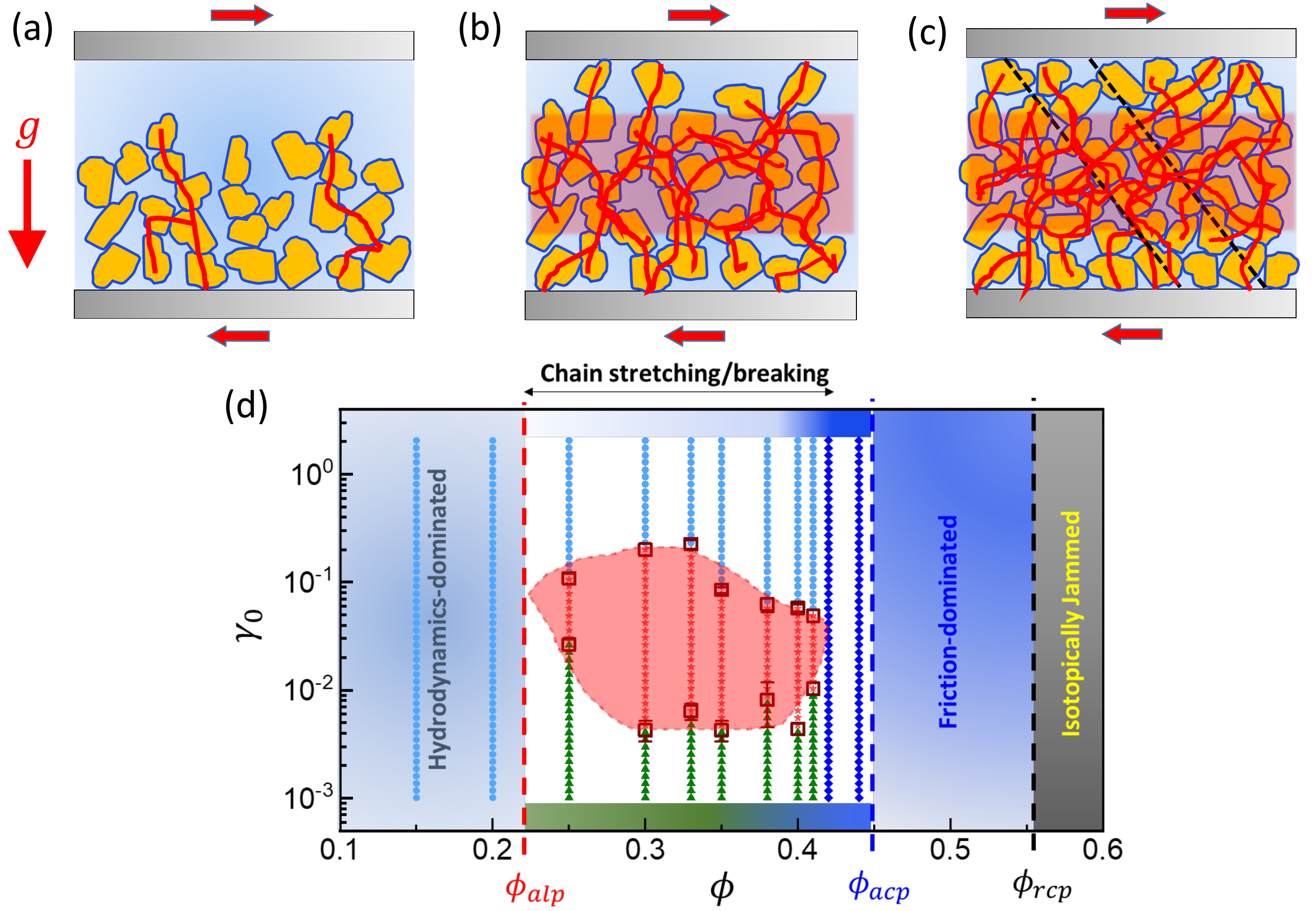}
    \caption{Schematics showing the configurations under gravity for different volume fractions ($\phi$) of CS in Paraffin oil during the rheology measurements. (a) For $\phi < \phi_{alp}$, the adhesive particle contacts (red solid lines) cannot span the entire gap due to settling. (b) $\phi_{alp} < \phi < \phi_{acp}$ contact networks span the gap. In this region, the system forms load bearing chains under extension but not along the compression direction. For even higher $\phi$ values (panel (c)), such compressive force chains (dashed black lines) stabilized by inter-particle friction can also form. The shaded region midway from both the plates (panels b and c) indicate the solid-like region inside the sample. These scenarios are summarized in a phase diagram in panel (d) in the parameter plane of $\phi$ and strain amplitude ($\gamma_0$). Different regions are marked based on the rheological response of the system. For $\phi < \phi_{alp}$ we get a viscous behaviour (light blue circles) over the entire range of $\gamma_0$. For $\phi_{alp} < \phi < \phi_{acp}$ we observe a non-monotonic strain response of $E_N$ related to the stretching and breaking of particle chains. The data points shown by green triangles indicate the region where $E_N \approx E_N^0$; red stars indicate the region where $E_N < E_N^0$ and strain stiffening is observed; light blue circles mark the region of $E_N > E_N^0$ where the system shows progressively larger dissipation. Such non-monotonicity disappears (dark blue diamonds) near $\phi \rightarrow \phi_{acp}$, where $E_N > E_N^0$ for all values of $\gamma_0$. $\phi = \phi_{rcp}$ indicates the random close packing for CS particles in the limit of hard-sphere repulsion. For $\phi \geq \phi_{rcp}$, the system is isotropically jammed. The critical volume fractions $\phi_{alp}$, $\phi_{acp}$ and $\phi_{rcp}$ are obtained from the particle-settling experiments. 
	}
    \label{F6}
    \end{center}
\end{figure}	
\\\\ 
As mentioned earlier, inter-particle interactions in dense suspensions are extremely complex and show an interesting dependence on the applied stress. In these systems the information about such interactions is encoded in the jamming packing fractions. To directly measure the jamming packing fractions for our system, we use particle settling experiments (Materials and Methods). In our case, the average volume fraction inside a stable settled bed ($\phi_{bed}$) for a particular forcing condition gives the corresponding jamming packing fraction. Essentially, $\phi_{bed}$ is the minimum particle volume fraction required to constrain the sliding/rolling motion of the particles against the stress acting on the bed. We show the typical images of stable settled beds formed under gravity (Fig. 4a) and centrifugation at 2000 rpm (Fig. 4b) for different interparticle interactions tuned by addition of surfactant (also see Fig. S6). We observe from Fig. 4a that for pure CS in oil system the height of the stable settled bed is significant with $\phi_{bed} \approx$ 0.22, indicating a highly porous particle arrangement inside the bed. This corresponds to the adhesive loose packing of the system ($\phi_{alp}$). However, once the adhesive interactions between the particles are reduced by introducing surfactant, the bed height decreases due to more compact particle organization. In Fig. 4c we show the variation of $\phi_{bed}$ as a function of surfactant concentration ($c$) for gravitational settling. With increasing $c$, $\phi_{bed}$ increases from $\phi_{alp}$ (for $c$ = 0) and saturates beyond $c >$ 0.03 \%. Similar trend of $\phi_{bed}$ vs $c$ is also observed under centrifugation (Fig. 4d), where the effective acceleration ($g_{eff}$) is much higher than the acceleration due to gravity ($g$) (Materials and Methods). We choose the value of $g_{eff}$ such that the resulting inter-particle stress scale is much higher than that probed in the rheology measurements (See S.I.). Under this condition, $\phi_{bed}$ (for $c$ = 0) gives the adhesive close packing ($\phi_{acp}$)  and the saturation at higher values of $c > 0.06$\% gives the random close packing ($\phi_{rcp}$) of the system in the limit of hard-sphere repulsion. We find $\phi_{acp} \approx$ 0.45 and $\phi_{rcp} \approx$ 0.55. The value of $\phi_{rcp}$ is close to that reported for CS particles with repulsive interactions \cite{Peters2016Apr}. We also confirm that the obtained values of $\phi_{acp}$ and $\phi_{rcp}$ are not sensitive to the starting volume fractions and different $g_{eff}$ values we use. 
\\\\
Our ability to tune the inter-particle interactions enables us to investigate the role of adhesion on the observed non-linear strain stiffening and the energy dissipation in the system. From rheological measurements, we find that increasing the amount of surfactant causes a dramatic reduction of the energy dissipation ($E$) in the system (Fig. S7). We show the variation of $E_N$ vs $\gamma_0$ for $\phi$ = 0.35, with different surfactant concentrations ($c$) in Fig. 5. Remarkably, we find that the non-monotonic behaviour of $E_N$ (Fig. 5) completely disappears with the addition of sufficient  amount of surfactant implying that strain stiffening also goes away under this condition (Fig. S8). This observation further confirms that adhesive interactions give rise to the non-linear strain stiffening in the system through the shear induced deformation of fractal particle clusters. Moreover, with increasing values of $c$, the velocity across the shearing gap also approaches a linear profile from a shear-banding one (Fig. S9).      
\\
\\Finally, we summarize our results for CS in oil system using a generalized phase diagram in Fig. 6. For lower values of $\phi$ below $\phi_{alp}$, gravitational settling forms a bed of particles near the bottom plate and a solvent layer is observed near the top plate of the rheometer, as shown by the schematic Fig. 6a. The contact networks between the particles can not span the entire gap between the shearing plates and the system behaviour remains viscous dominated over the entire range of strain amplitudes (Fig. 6d). However, for $\phi$ values larger than $\phi_{alp}$, the settled particle bed can span the entire shear-gap (schematic Fig. 6b). The system develops a finite yield stress like a soft visco-elastic solid. In the regime $\phi > \phi_{alp}$ (but well below $\phi_{acp}$), the system shows significant resistance in response to applied strain due to the stretching of adhesive contacts, however, $\phi$ is still low enough to not support system spanning force chains along the compression direction. Here, the system transforms from a quasi-linear visco-elastic solid to a strain-stiffening solid ($E_N < E_N^0$) and finally to a viscous/plastic material with increasing $\gamma_0$, as shown in Fig. 6d. Similar strain stiffening at intermediate strain values has also been observed for colloidal gels formed by Brownian particles \cite{vanDoorn2018May}. We note that the strain stiffening does not take place for $\gamma_0 < 0.003$ (Fig. 6d). This indicates that significant deformation of the fractal clusters are required for the observed strain stiffening response. Interestingly, the strain stiffening disappears for $\phi$ values close to $\phi_{acp}$. For such high volume fractions, the system spanning force chains can also form along the compression direction (Fig. 6c) \cite{Richards2020Mar, Peters2016Apr}. Thus, strain deformations give rise to considerable frictional interaction between the particles. This results in an enhanced dissipation masking the strain stiffening response. This explains the fact that despite significant adhesive interactions, $E_N$ remains higher than $E_N^0$ in this regime. We also observe that close to $\phi_{acp}$ sample mixing becomes extremely difficult and sample appears almost dry. Due to this we can not experimentally probe the regime $\phi_{acp} < \phi < \phi_{rcp}$ (Fig. 6d). We want to reemphasize that the complex dynamics originating from the stretching/breaking of adhesive contacts and formation of frictional contacts take place near the shearing boundaries, while the bulk of the sample moves like a solid-plug as pointed out in Fig. 6b and 6c. This is a major new addition to the recent physical picture of the flow behaviour in similar systems \cite{Richards2020Mar}. 

\section{Conclusion}
We study yielding and energy dissipation in granular suspensions of adhesive frictional particles. We find that the normalized energy dissipation $E_N$ shows an interesting non-monotonic dependence on both applied strain and volume fraction. We show that such non-monotonic behaviour is intimately linked to the interplay between interparticle adhesion and friction in the system. From optical imaging, we observe strain localization and random spatio-temporal fluctuations in local velocity gradients. Using stroboscopic image sampling, we demonstrate a direct correlation between such fluctuations and irreversible particle rearrangements.

A salient feature of our study is that we directly measure the critical jamming packing fractions using particle-settling experiments. Thus, besides the inter-particle interactions, particle shapes are automatically taken into account. Particle shape is an important parameter in constraint rheology models that can significantly affect the inter-particle rolling degrees of freedom. This is particularly desirable for amorphous systems where the particle shape parameter is difficult to include in numerical simulations. Remarkably, we show that the critical jamming packing fractions, estimated from the particle-settling experiments, successfully capture the essential physics behind the different flow regimes observed over a wide range of volume fractions and applied strain amplitudes. 
   
The coexisting solid- and fluid-like regions imply that for adhesive systems, stress/strain induced breaking of interparticle adhesive contacts does not happen uniformly across the entire system, even beyond yielding. This observation underscores the importance of strain localization and puts additional constraints on the recently proposed models \cite{guy2018constraint, Richards2020Mar} for yielding in adhesive granular systems. 
 
The non-linear strain stiffening observed for intermediate applied strain values is reminiscent of  similar phenomena in semiflexible biopolymer networks \cite{Storm2005May} and colloidal gels forming strand-like structures \cite{vanDoorn2018May}. Using fluorescently labeled CS particles, we indeed observe system-spanning network-like connected structures formed by fractal clusters of adhesive particles. Thus, for intermediate packing fractions, the strain stiffening can take place due to the stretching of these contacts for moderate strain values, but larger applied strains can break the contacts giving rise to enhanced plasticity and dissipation. Addition of surfactant inhibits the adhesive interactions and thus disrupts the formation of such system spanning connected structures. Consequently, no strain stiffening is observed. Importantly, the presence of surfactant molecules on the particle surface not only reduces the interparticle adhesion, but can also significantly modify the interparticle friction coefficients. However, quantifying such effects in our system remains an interesting future direction to explore.  

Our study provides a complete picture of flow and yielding behaviour in dense granular suspensions of adhesive amorphous particles and can have important implications for both theoretical as well as experimental studies in future.  

\section{Materials and Methods}
	For our measurements, the samples are made by dispersing Cornstarch (CS) particles (Sigma Aldrich) in Paraffin oil (SDFCL) at different volume fractions $\phi$ ranging from 0.15 to 0.44. To prepare the samples for $\phi \leq$ 0.3, CS powder is gradually added to the oil and mixed thoroughly using a magnetic stirrer. For samples with higher $\phi$ values, a combination of hand mixing and magnetic stirring is employed to ensure homogeneity of the sample. For making the suspensions with repulsive inter-particle interactions, we disperse CS in either water or in a water-glycerol mixture having viscosity matched with the paraffin oil at room temperature (25 $^{o}$C). The suspensions prepared using water-glycerol mixture are sonicated for 20 to 30 minutes after hand mixing to ensure homogeneity. To tune the inter-particle adhesive interactions, we use a nonionic surfactant Span\textsuperscript{\textregistered} 60 (Sigma Aldrich). To make the CS suspensions in oil with added surfactant, we first weigh out the required amount of surfactant (in powder form) and then crush it in a mortar pestle to get rid of big clusters, if any. Next, CS particles are added and dry mixed thoroughly with the surfactant. After that, the oil is added to the dry mixture and then the sample is mixed well in the mortar pestle followed by mixing with a magnetic stirrer till homogenized. All the samples are degassed overnight under vacuum at room temperature in a desiccator (Borosil)/vacuum oven (Allied Scientific) before rheology/particle settling experiments.  
	\\For particle settling experiments, the degassed samples are first transferred to 15 mL graduated Falcon tubes very gently. Next, the tubes are either kept in a vertical position and left undisturbed for 3 weeks (for settling under gravity) or, centrifuged (Remi, R-8C BL) in a swinging-bucket type holder with rotation speed varying between 1000 - 2500 rpm for 2 hours. For this range of rotation speeds and considering slight sample to sample variation in bed heights, we find that the average acceleration approximately varies between 114$g$ - 711$g$ ($g$: acceleration due to gravity). We confirm that in all cases the waiting time is sufficient to get a stable bed formation.
	\\For confocal microscopy, we use the fluorescent ink extracted from Faber Castell Textliner Supefluorescent markers as the dye. The excitation and emission wavelengths are 435 nm and 570 nm, respectively, as obtained from the absorption (using UV-Visible Spectroscopy, Perkin Elmer Lambda 35) and the emission spectrum (using Photoluminescence spectroscopy, Horiba Jobin Yvon – Edison, NJ USA). Sample preparation for confocal microscopy involves adding the CS particles to a petri dish containing the dye dissolved in ethanol. The solvent is then evaporated at room temperature to get the dyed CS particles. The particles are dried further in a vacuum oven. For making dense suspensions using these dyed CS particles dispersed in oil (with/without surfactant) we follow a similar mixing protocol as described above. For confocal imaging we use a Leica DMI6000 microscope and  a confocal scanner (Sp8 Germany). The z-stack images are obtained with a z-spacing of 0.68 $\mu$m.
	\\Rheological measurements are performed using a stress controlled rheometer (MCR-702 Anton Paar, Austria). We use a cone-plate geometry with diameter of 50 mm and cone angle of 2\textsuperscript{o}. Both the surfaces are sand-blasted to minimize slippage at the sample boundaries. We use Large amplitude oscillatory shear (LAOS) protocol for all our rheological measurements in the separate motor transducer (SMT) mode, where the bottom plate is moving and the top plate is held stationary. LAOS measurements are done at an angular frequency ($\omega$) of 0.1 rad/s and the strain amplitude is varied from 0.1\% to 200\% with the measurement time per point set by the device (which is found to be roughly over the duration of 3 cycles).  For the rheological measurements on repulsive systems (using water-glycerol mixture as solvent), an in-house built humidity chamber is used to prevent solvent evaporation. 
	\\The in-situ imaging is done using a Lumenera Lt545R camera fitted with a 5X Mitutoyo objective. For image analysis, the images are taken at discrete values of $\gamma_0$ with frame rate varying between 1 Hz - 40 Hz.
	
\section{Authors Contributions}
S.C. and S.M. designed the research, S.C. and S.N. performed the experiments, S.C., S.N. and S.M. analyzed the data. S.C and S.M. wrote the manuscript.

\section{Conflict of interests}
There is no conflict to declare.

\section{Acknowledgments}
S.M. thanks SERB (under DST, Govt. of India) for a Ramanujan Fellowship. We acknowledge Ivo Peters for developing the Matlab codes used for PIV analysis, K M Yatheendran for help with the SEM imaging and RRI workshop facility for machining the humidity chamber. We thank Pinaki Chaudhuri for helpful discussion.
	

\newpage
\section{Movie Descriptions}

\textbf{Movie 1:}\,\,\,\,In this movie, we show an animation of the reconstructed 3D images (obtained from the maximum intensity projection for each image in the z-stack) of CS in oil system without (left panel) and with 0.2 \% surfactant (right panel) under 20X magnification. In both the cases, the initial volume fraction of CS is 0.05. The z-stack images were captured using a confocal microscope with a z-spacing of 0.68 $\mu$m once the particles are settled to form a stable bed. Since the intensity of the images gets reduced with increasing depth due to the opaque nature of the particles, gamma and brightness/contrast correction was applied to the images before the 3D reconstruction. It can be seen that the thickness of the 3D volume is more for the adhesive system (without surfactant)  as compared to that for the repulsive system (with added surfactant). This confirms our hypothesis that once the interparticle adhesive interaction is removed, the particle strands can no longer sustain their weight and hence collapse to form a more uniformly distributed bed. The 2D distribution of the particles in the x-y plane (as seen from the starting images in the animation) shows a very uniform distribution  in the repulsive case (with surfactant) as compared to the adhesive case (without surfactant) where clear particle aggregation and void formation can be observed.  
\newline
\newline
\textbf{Movie 2:}\,\,\,\,In this movie, we show an animation of the reconstructed 3D images of CS in oil system without (left panel) and with surfactant (right panel) under 40X magnification. Other parameters are same as Movie 1. Once again the thickness of the 3D volume is more in the case without surfactant (the adhesive system) as compared to the surfactant case (the repulsive system). The 2D image of the particle distribution in x-y plane with the higher magnification shows the structural heterogeneity in the adhesive case even more clearly than that observed with 20X.  

\section{Estimation of average stress scale inside the settled bed}
To estimate the average interparticle stress inside a stable settled bed, we consider the total stress acting on a CS particle at the mid-point (point `O') of the settled bed (along the length of the tube). Here, we assume a uniform distribution of CS particles inside the settled bed. The total stress on a particle at `O' has two contributions: stress acting on a particle at this point assuming the particle to be isolated and the stress due to part of the bed above it (for gravitational settling) or, the part towards the axis of rotation (for settling under centrifugation). \\ Considering a typical case of centrifugation, the distance of the mid-point of bed from the axis of rotation is $R = 15.17$ cm. The length of half of the settled bed is $l = 1.77$ cm. The distance of the center of mass of the part closer to the axis of rotation compared to the point `O' is $R_1 = R - l/2 = 14.3$ cm. We use $R_{CS}$ = 7.5 $\mu$m (Refer S1) as the average particle radius, and the rotation speed $\omega \approx$ 209 rad/s (converted from 2000 rpm). The density of CS and paraffin oil (PO) is 1.6 g/cc and 0.89 g/cc respectively. The volume fraction of the starting sample is 0.2, however, after centrifugation the volume fraction inside the bed is found to be $\phi_{bed} \approx$ 0.44. 
\\So, the stress on the free particle is ${\sigma_{isolated}} = \frac{4}{3} R_{CS} (\rho_{CS} - \rho_{PO})\,\omega^2 R $ and that due to half of the bed above is ${\sigma_{bed}} = l\,(\rho_{CS} - \rho_{PO})\,\omega^2 R_1 \phi_{bed}$. The total stress value we get is 35 kPa which is much higher than the maximum stress we observe in our rheological measurements ($\approx$ 150 Pa) for the sample with the highest volume fraction ($\phi$ = 0.44). 
\\Similarly for gravitational settling, the stress on the isolated particle is $\sigma_{isolated} = \frac{4}{3} R_{CS}(\rho_{CS} - \rho_{PO}) g$ and stress from the half of the bed on top is $\sigma_{gravity} = l (\rho_{CS} - \rho_{PO}) g$, the sum of which comes out to be 123 Pa (we use $g = 9.8 m/{s^2}$). 
\\In both the cases of settling under gravity and centrifugation, we find that the contribution from the isolated particle is negligible in comparison to the contribution from the part of the settled bed that comes into play.

\section{Calculation of fractal dimension inside the settled bed}
As mentioned in the main text, due to the adhesive interaction between the particles (CS in PO), we get a stable settled bed that can self-support it's own weight even for low volume fractions well below $\phi_{rcp}$. This implies the formation of a porous/fractal structure inside the bed as also observed from the confocal imaging (Movie 1 and Movie 2). The fractal dimension of the system is dependent on the initial volume fraction of particles for $\phi > \phi_{alp}$. Due to the opaque nature of CS, the 3-D fractal dimension is very hard to calculate from confocal imaging, particularly for packing fraction much larger than $\phi_{alp}$. We use particle settling experiments to estimate the 3-d fractal dimension of the system. We assume that $\phi_{rcp}$ being the most efficient random packing in the system, the mass contained in a sphere of radius $R'$, $M(R') = m(R'/a)^3$, where $m$ and $a$ represents the mass and radius of a single CS particle. However, the mass contained within a radius $R$ in presence of adhesive interaction, $M(R) = m(R/a)^{d_f}$ with $d_f < 3$ representing the porous structure inside the bed. With reference to the particle settling experiments shown in Fig. S3a and S3b, $V_b$ represents the volume of the settled bed in presence of adhesive interaction and $V'_b$ represents the volume occupied by the same system assuming the volume fraction inside the bed to be $\phi_{rcp}$. Since, the total mass of the particles contained in $V_b$ and $V'_b$ is the same, we get, $M(R) = m(R/a)^{d_f}$ and $M(R') = m(R'/a)^3$ where $R$ and $R'$ represents the radii corresponding to the spherical volumes equivalent to $V_b$ and $V'_b$, respectively. Thus, ${V_b}= {4/3}\,\pi R^3 $ and ${V'_b}= {4/3}\,\pi R'^3 $. Also, ${\phi}{V_b} = {\phi_{rcp}}{V'_b}$. Using these, we get the expression for the fractal dimension as $d_f = 3\,[\frac{\log(R'/a)}{\log(R/a)}]$. Putting the values for $R'$ and $R$ we get $d_f$ as a function of $\phi$ as shown in Fig. S3c. 

\newpage
\section{Supplementary figures}
\begin{figure} [h]
    \begin{center}
    \includegraphics[height = 7 cm]{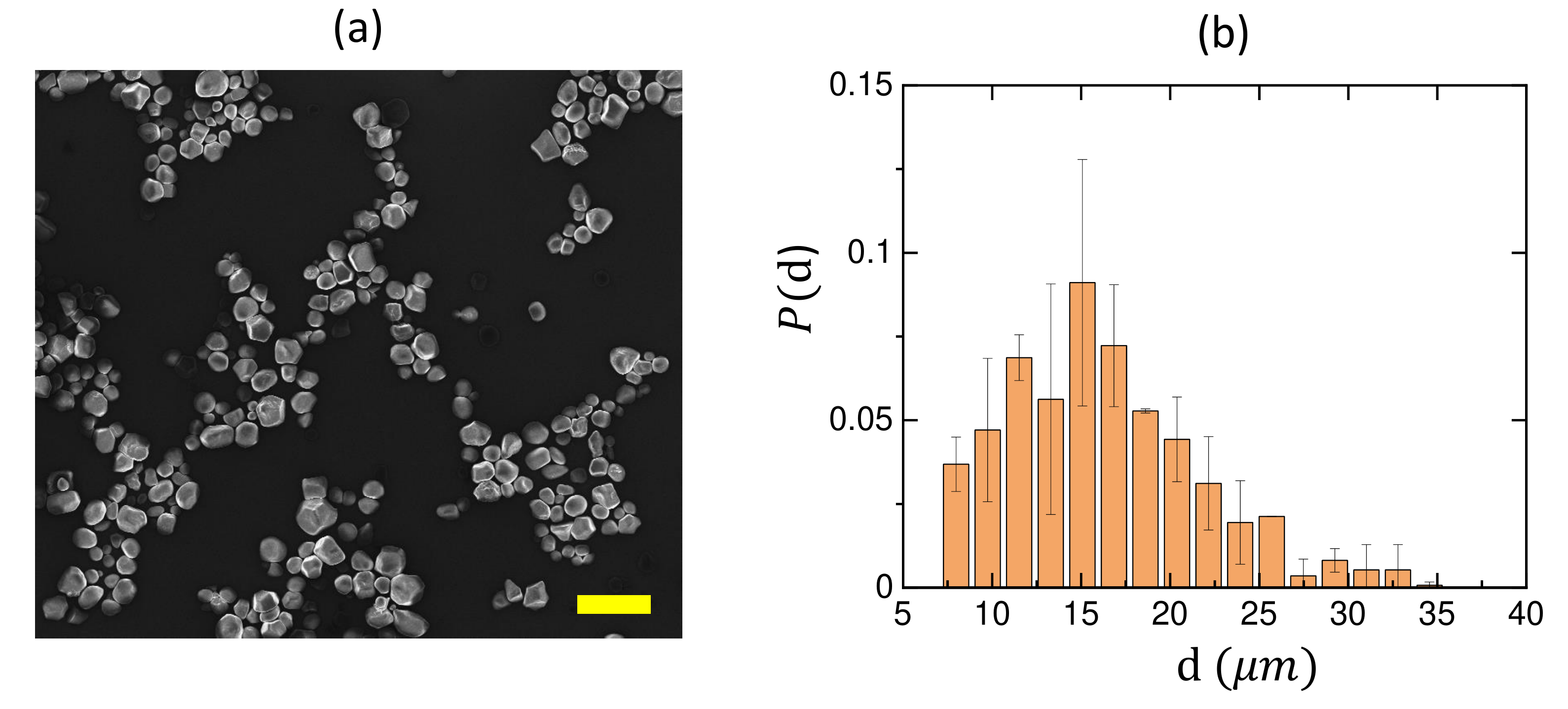}
		\renewcommand{\thefigure}{S1}
		\vspace*{-3mm}
    \caption{(a) A typical SEM image of dry CS particles. The scalebar is 40 $\mu$m. (b) Histogram shows the particle size distribution with mean diameter d = 15 $\mu$m and polydispersity (standard deviation / mean) = 0.3 $\mu$m. Error bars indicate the standard deviation of the four different histograms obtained from different SEM images.}
    \label{S1}
    \end{center}
\end{figure}
\begin{figure} [h]
    \begin{center}
    \includegraphics[height = 7.5 cm]{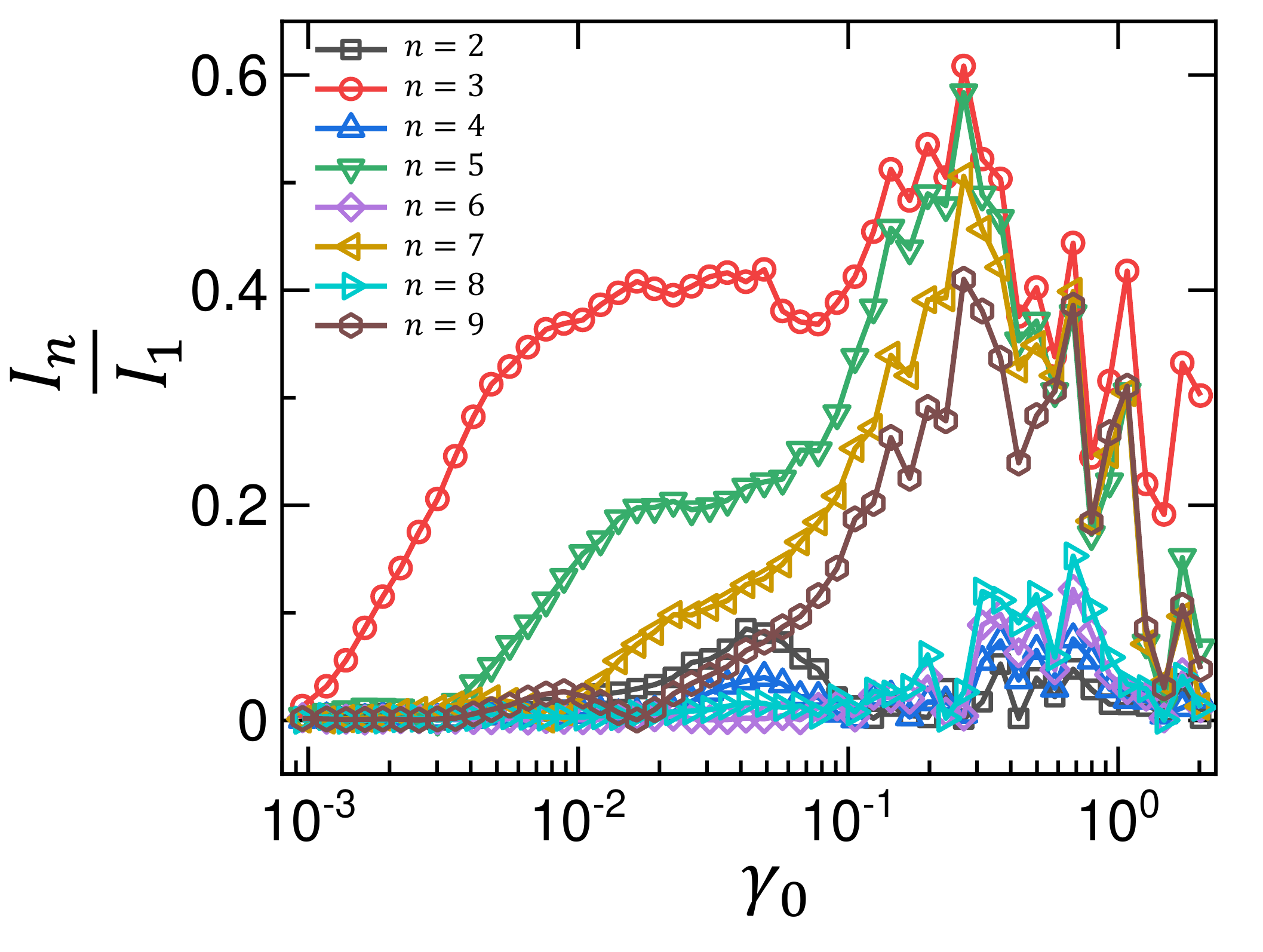}
		\renewcommand{\thefigure}{S2}
		\vspace*{-3mm}
    \caption{Relative contribution of the higher harmonics with respect to the fundamental ($I_n/I_1$) as a function of $\gamma_0$ for volume fraction ($\phi$) $=$ 0.4 (corresponding to the data shown in Fig. 1(a)). Here, $I_n/I_1 = |G'_n/G'|$.}
    \label{S2}
    \end{center}
\end{figure}
\begin{figure}[h]
    \begin{center}
    \includegraphics[height = 7.2 cm]{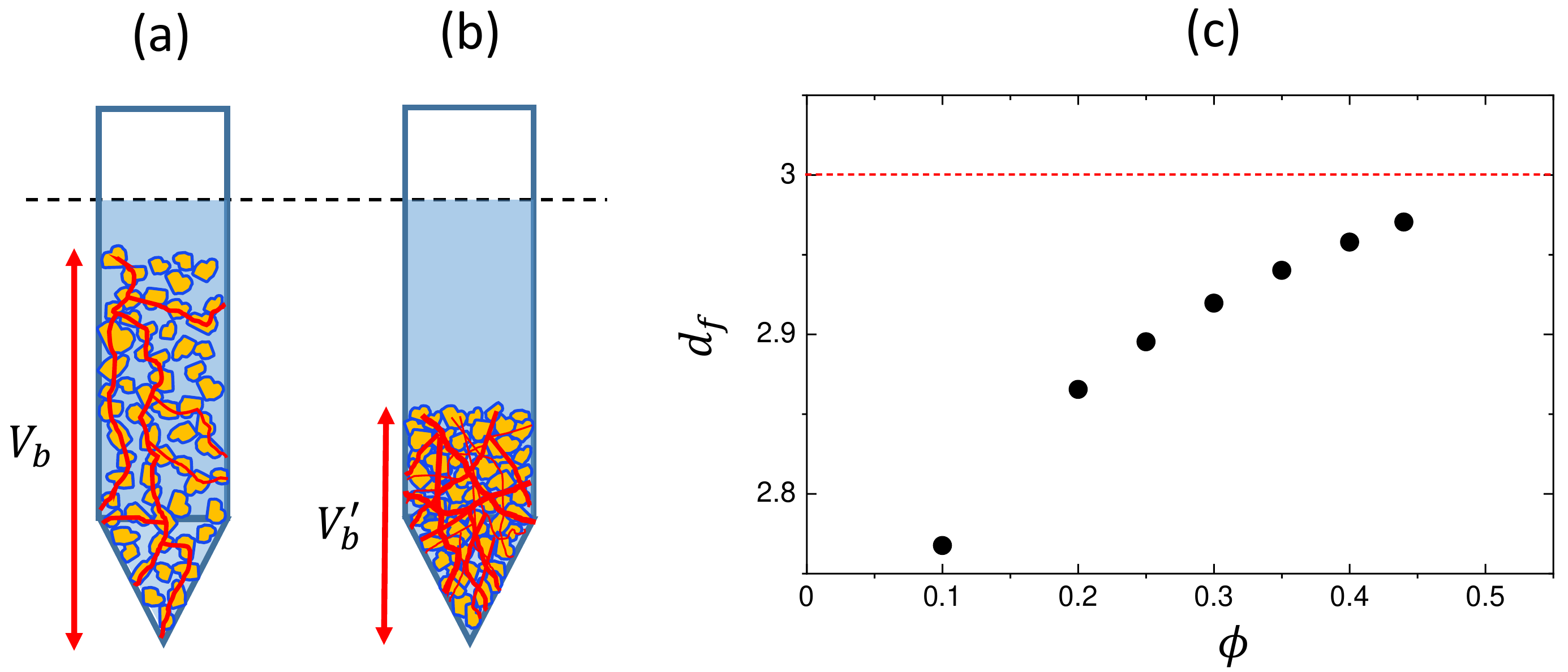}
		\renewcommand{\thefigure}{S3}
		\vspace*{-3mm}
    \caption{(a) Typical porous arrangement of the particles inside a stable settled bed formed under gravity for $\phi < \phi_{alp}$ in case of CS in paraffin oil. The system spanning adhesive contacts stabilizing the bed are marked with red lines. (b) The same system in the absence of adhesive interaction and under high stress becomes more compact with $\phi \rightarrow \phi_{rcp}$. Here, the system spanning particle contacts (red lines) are due to hard sphere interaction between the particles. (c) Variation of 3-D fractal dimension ($D_f$) as a function of initial volume fraction ($\phi$) obtained from the particle settling experiments. Detailed calculation is given in S.I. text.}
    \label{S3}
    \end{center}
\end{figure}

\begin{figure}[h]
    \begin{center}
    \includegraphics[height = 7.5 cm]{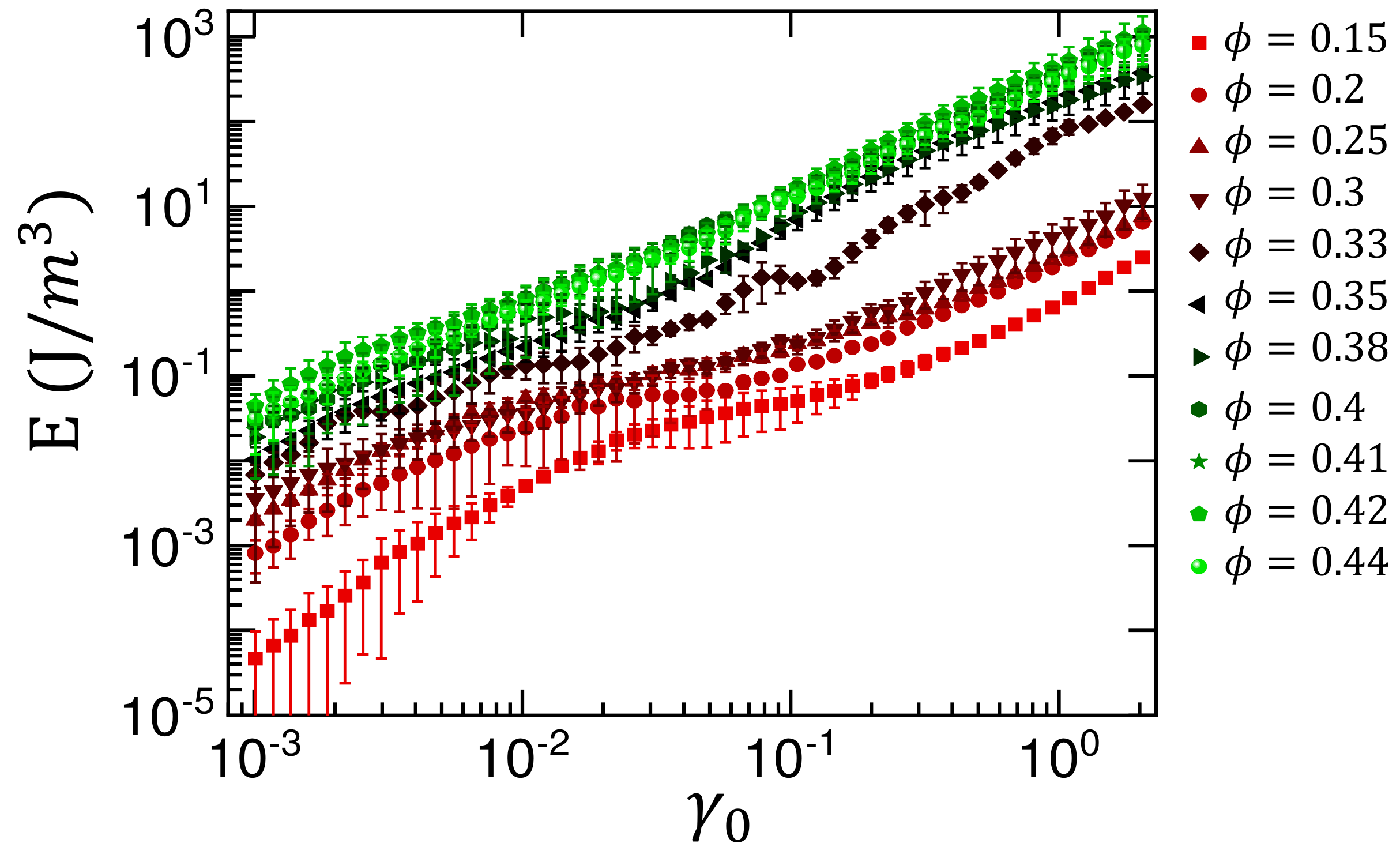}
		\renewcommand{\thefigure}{S4}
    \caption{Variation of dissipated energy density ($E$) as a function of $\gamma_0$ for CS particles dispersed in oil for $\phi$ ranging from 0.15 to 0.44.}
    \label{S4}
    \end{center}
\end{figure}
\begin{figure}[h]
    \begin{center}
    \includegraphics[height = 10 cm]{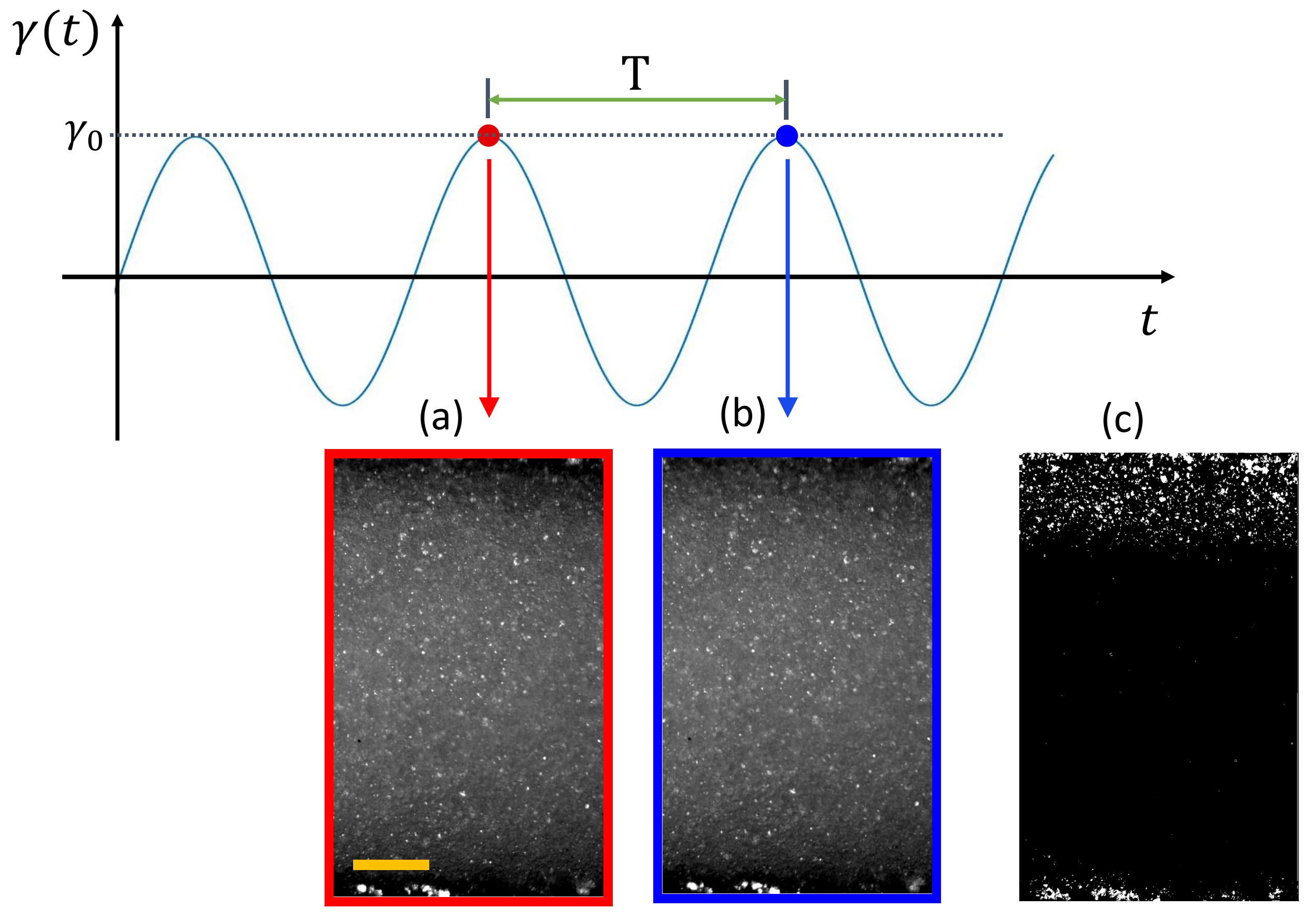}
		\renewcommand{\thefigure}{S5}
		\vspace*{-3mm}
    \caption{Top panel shows a typical sinusoidal strain input from the LAOS measurements. Points indicated in red and blue are at the same part of the cycles (here chosen at the turning points) but, separated by one time period (T) as indicated. The panels (a) and (b)  show the images taken at the shown points (image frame colours correspond to the points on the sinusoidal curve). Image (c) represents the stroboscopic image which is obtained by taking a difference between the images shown in (a) and (b). Scale bar represents 150 $\mu$m.}
    \label{S5}
    \end{center}
\end{figure}
\begin{figure}[h]
    \begin{center}
    \includegraphics[height = 16 cm]{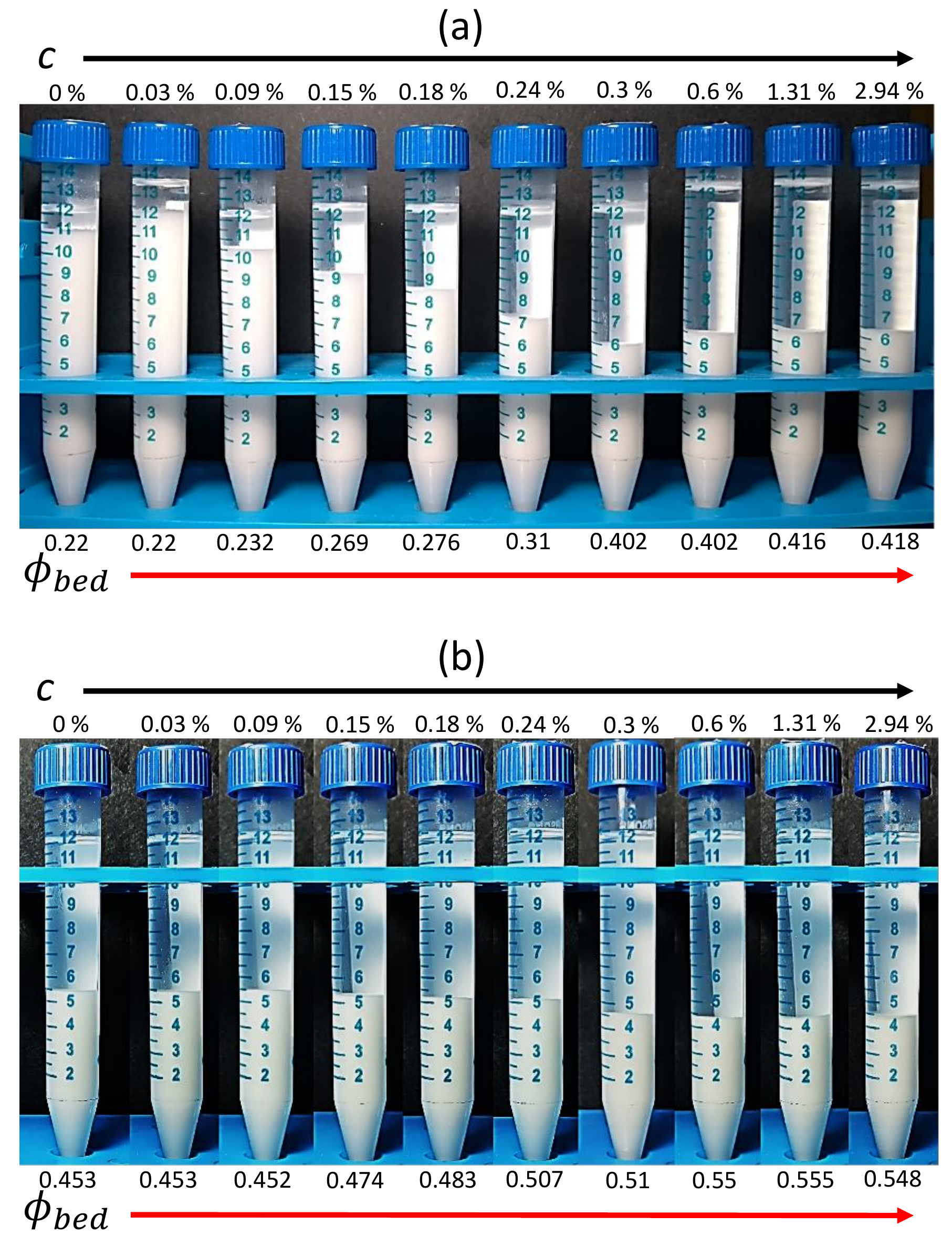}
		\renewcommand{\thefigure}{S6}
		\vspace*{-3mm}
    \caption{Images of the settled beds as function of surfactant concentration ($c$) for gravitational settling (panel (a)) and centrifugation at 2000 rpm (panel (b)). The particle volume fraction inside the beds ($\phi_{bed}$) are also indicated. In all cases the initial volume fraction ($\phi$) of the suspensions is 0.2.}
    \label{S6}
    \end{center}
\end{figure}
\begin{figure}[h]
    \begin{center}
    \includegraphics[height = 8.5 cm]{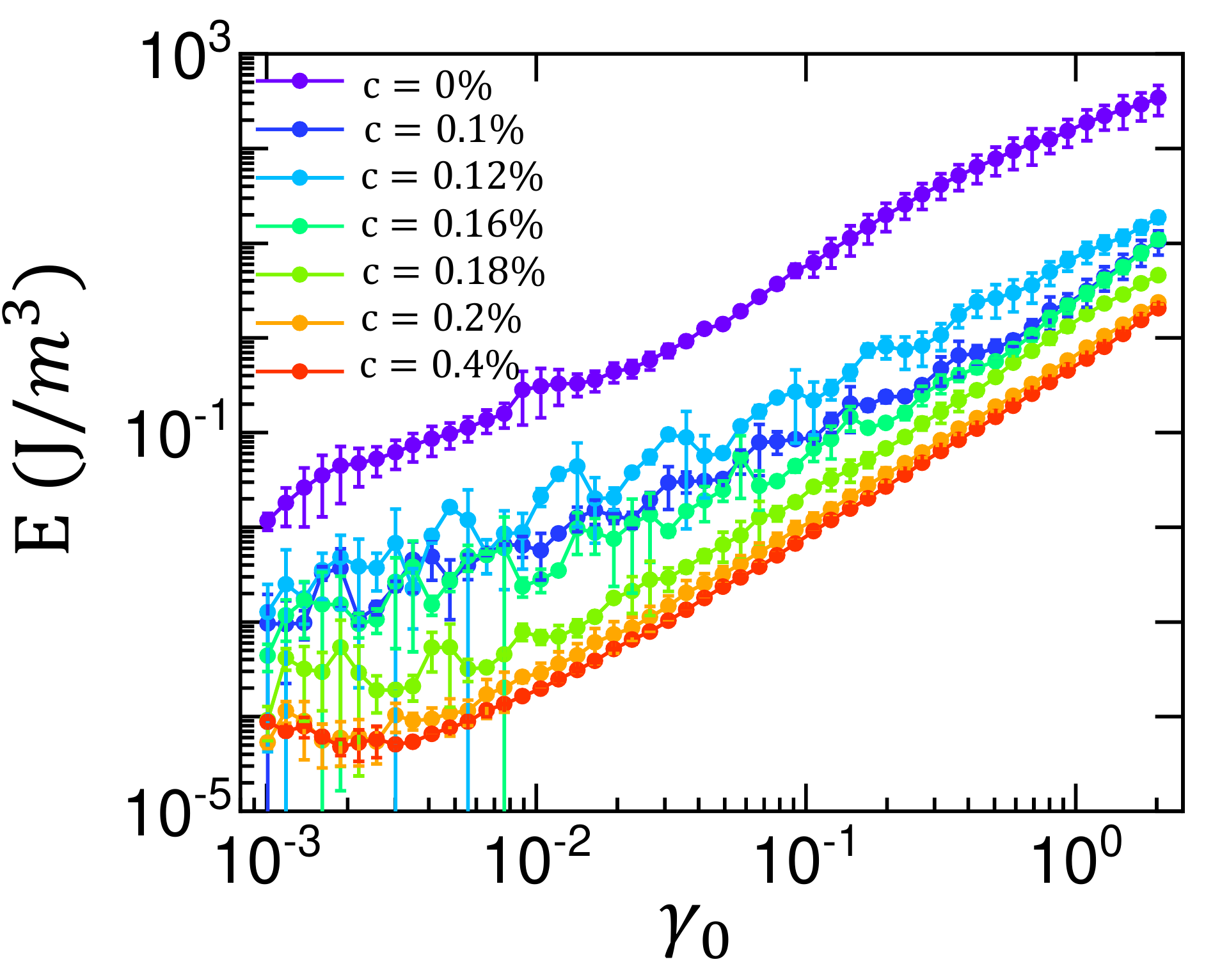}
		\renewcommand{\thefigure}{S7}
		\vspace*{-3mm}
    \caption{Variation of dissipated energy density ($E$) as a function of $\gamma_0$ for $\phi$ = 0.35 with increasing concentration of surfactant ($c$). }
    \label{S7}
    \end{center}
\end{figure}
\begin{figure}[h]
    \begin{center}
    \includegraphics[height = 7 cm]{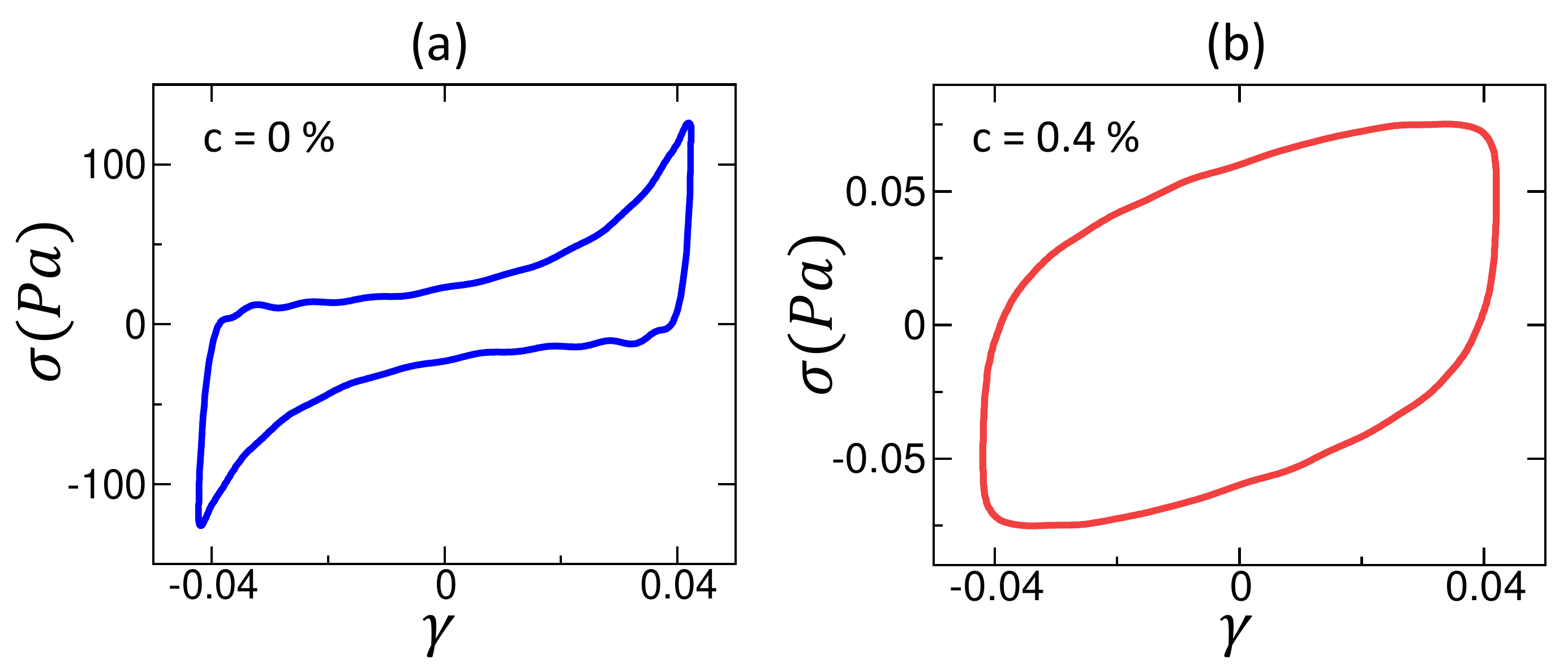}
		\renewcommand{\thefigure}{S8}
		\vspace*{-3mm}
    \caption{Elastic Lissajous plot at an intermediate value of $\gamma_0$ during the LAOS measurement for $\phi$ = 0.4 without surfactant (panel (a)) and with surfactant concentration c = 0.4 \% (right panel (b)). We see that with the addition of sufficient amount of surfactant the stress values dramatically drop and also the strain stiffening behaviour completely disappears.   
		}
    \label{S8}
    \end{center}
\end{figure}
\begin{figure}[h]
    \begin{center}
    \includegraphics[height = 7.2 cm]{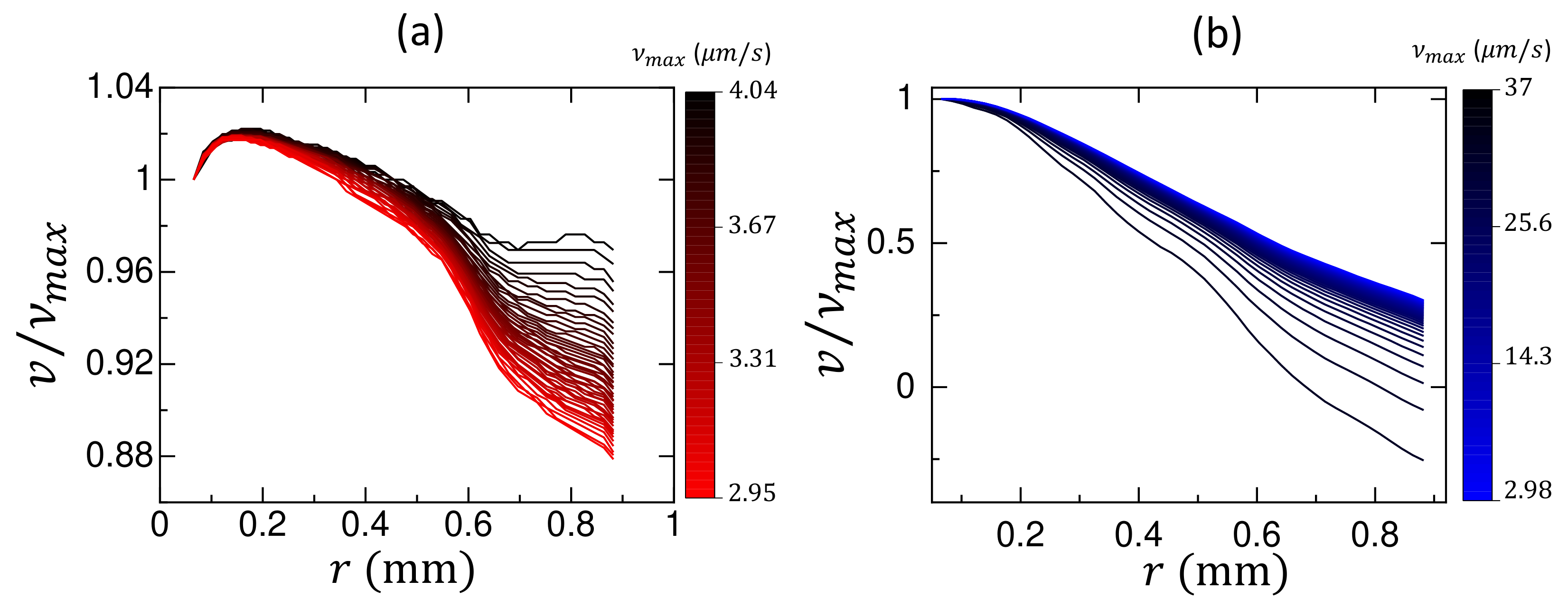}
		\renewcommand{\thefigure}{S9}
		\vspace*{-3mm}
    \caption{Velocity profiles across the gap between the two plates normalized by the instantaneous maximum velocity of the sample (near the moving plate beyond the plate roughness) for sample volume fraction $\phi$ = 0.4 with surfactant concentration c = 0.4 \% at $\gamma_0$ = 0.1 (left panel) and $\gamma_0$ = 0.5 (right panel). The colour gradient indicates the instantaneous maximum velocity in the sample (approximately equal to the plate velocity at that instant) during an oscillatory applied strain cycle.    
		}
    \label{S9}
    \end{center}
\end{figure}

\begin{thebibliography}{51}
\expandafter\ifx\csname natexlab\endcsname\relax\def\natexlab#1{#1}\fi
\expandafter\ifx\csname bibnamefont\endcsname\relax
  \def\bibnamefont#1{#1}\fi
\expandafter\ifx\csname bibfnamefont\endcsname\relax
  \def\bibfnamefont#1{#1}\fi
\expandafter\ifx\csname citenamefont\endcsname\relax
  \def\citenamefont#1{#1}\fi
\expandafter\ifx\csname url\endcsname\relax
  \def\url#1{\texttt{#1}}\fi
\expandafter\ifx\csname urlprefix\endcsname\relax\def\urlprefix{URL }\fi
\providecommand{\bibinfo}[2]{#2}
\providecommand{\eprint}[2][]{\url{#2}}

\bibitem[{\citenamefont{Falk and Langer}(2011)}]{Falk2011Feb}
\bibinfo{author}{\bibfnamefont{M.~L.} \bibnamefont{Falk}} \bibnamefont{and}
  \bibinfo{author}{\bibfnamefont{J.~S.} \bibnamefont{Langer}},
  \bibinfo{journal}{Annu. Rev. Condens. Matter Phys.}
  \textbf{\bibinfo{volume}{2}}, \bibinfo{pages}{353} (\bibinfo{year}{2011}).

\bibitem[{\citenamefont{Berthier et~al.}(2011)\citenamefont{Berthier, Biroli,
  Bouchaud, Cipelletti, and van Saarloos}}]{Berthier2011Jul}
\bibinfo{author}{\bibfnamefont{L.}~\bibnamefont{Berthier}},
  \bibinfo{author}{\bibfnamefont{G.}~\bibnamefont{Biroli}},
  \bibinfo{author}{\bibfnamefont{J.-P.} \bibnamefont{Bouchaud}},
  \bibinfo{author}{\bibfnamefont{L.}~\bibnamefont{Cipelletti}},
  \bibnamefont{and} \bibinfo{author}{\bibfnamefont{W.}~\bibnamefont{van
  Saarloos}}, \emph{\bibinfo{title}{{Dynamical Heterogeneities in Glasses,
  Colloids, and Granular Media: 150 (International Series of Monographs on
  Physics)}}} (\bibinfo{year}{2011}).

\bibitem[{\citenamefont{Bonn et~al.}(2017)\citenamefont{Bonn, Denn, Berthier,
  Divoux, and Manneville}}]{Bonn2017Aug}
\bibinfo{author}{\bibfnamefont{D.}~\bibnamefont{Bonn}},
  \bibinfo{author}{\bibfnamefont{M.~M.} \bibnamefont{Denn}},
  \bibinfo{author}{\bibfnamefont{L.}~\bibnamefont{Berthier}},
  \bibinfo{author}{\bibfnamefont{T.}~\bibnamefont{Divoux}}, \bibnamefont{and}
  \bibinfo{author}{\bibfnamefont{S.}~\bibnamefont{Manneville}},
  \bibinfo{journal}{Rev. Mod. Phys.} \textbf{\bibinfo{volume}{89}},
  \bibinfo{pages}{035005} (\bibinfo{year}{2017}).

\bibitem[{\citenamefont{Coussot}(2014)}]{Coussot2014Sep}
\bibinfo{author}{\bibfnamefont{P.}~\bibnamefont{Coussot}}, \bibinfo{journal}{J.
  Non-Newtonian Fluid Mech.} \textbf{\bibinfo{volume}{211}},
  \bibinfo{pages}{31} (\bibinfo{year}{2014}).

\bibitem[{\citenamefont{van~der Vaart et~al.}(2013)\citenamefont{van~der Vaart,
  Rahmani, Zargar, Hu, Bonn, and Schall}}]{vanderVaart2013Jun}
\bibinfo{author}{\bibfnamefont{K.}~\bibnamefont{van~der Vaart}},
  \bibinfo{author}{\bibfnamefont{Y.}~\bibnamefont{Rahmani}},
  \bibinfo{author}{\bibfnamefont{R.}~\bibnamefont{Zargar}},
  \bibinfo{author}{\bibfnamefont{Z.}~\bibnamefont{Hu}},
  \bibinfo{author}{\bibfnamefont{D.}~\bibnamefont{Bonn}}, \bibnamefont{and}
  \bibinfo{author}{\bibfnamefont{P.}~\bibnamefont{Schall}},
  \bibinfo{journal}{J. Rheol.} \textbf{\bibinfo{volume}{57}},
  \bibinfo{pages}{1195} (\bibinfo{year}{2013}).

\bibitem[{\citenamefont{Patinet et~al.}(2016)\citenamefont{Patinet,
  Vandembroucq, and Falk}}]{patinet2016connecting}
\bibinfo{author}{\bibfnamefont{S.}~\bibnamefont{Patinet}},
  \bibinfo{author}{\bibfnamefont{D.}~\bibnamefont{Vandembroucq}},
  \bibnamefont{and} \bibinfo{author}{\bibfnamefont{M.~L.} \bibnamefont{Falk}},
  \bibinfo{journal}{Physical review letters} \textbf{\bibinfo{volume}{117}},
  \bibinfo{pages}{045501} (\bibinfo{year}{2016}).

\bibitem[{\citenamefont{Richard et~al.}(2020)\citenamefont{Richard, Ozawa,
  Patinet, Stanifer, Shang, Ridout, Xu, Zhang, Morse, Barrat
  et~al.}}]{Richard2020Nov}
\bibinfo{author}{\bibfnamefont{D.}~\bibnamefont{Richard}},
  \bibinfo{author}{\bibfnamefont{M.}~\bibnamefont{Ozawa}},
  \bibinfo{author}{\bibfnamefont{S.}~\bibnamefont{Patinet}},
  \bibinfo{author}{\bibfnamefont{E.}~\bibnamefont{Stanifer}},
  \bibinfo{author}{\bibfnamefont{B.}~\bibnamefont{Shang}},
  \bibinfo{author}{\bibfnamefont{S.~A.} \bibnamefont{Ridout}},
  \bibinfo{author}{\bibfnamefont{B.}~\bibnamefont{Xu}},
  \bibinfo{author}{\bibfnamefont{G.}~\bibnamefont{Zhang}},
  \bibinfo{author}{\bibfnamefont{P.~K.} \bibnamefont{Morse}},
  \bibinfo{author}{\bibfnamefont{J.-L.} \bibnamefont{Barrat}},
  \bibnamefont{et~al.}, \bibinfo{journal}{Phys. Rev. Mater.}
  \textbf{\bibinfo{volume}{4}}, \bibinfo{pages}{113609} (\bibinfo{year}{2020}).

\bibitem[{\citenamefont{Lin et~al.}(2014)\citenamefont{Lin, Lerner, Rosso, and
  Wyart}}]{Lin2014Oct}
\bibinfo{author}{\bibfnamefont{J.}~\bibnamefont{Lin}},
  \bibinfo{author}{\bibfnamefont{E.}~\bibnamefont{Lerner}},
  \bibinfo{author}{\bibfnamefont{A.}~\bibnamefont{Rosso}}, \bibnamefont{and}
  \bibinfo{author}{\bibfnamefont{M.}~\bibnamefont{Wyart}},
  \bibinfo{journal}{Proc. Natl. Acad. Sci. U.S.A.}
  \textbf{\bibinfo{volume}{111}}, \bibinfo{pages}{14382}
  (\bibinfo{year}{2014}).

\bibitem[{\citenamefont{Koumakis and Petekidis}(2011)}]{koumakis2011two}
\bibinfo{author}{\bibfnamefont{N.}~\bibnamefont{Koumakis}} \bibnamefont{and}
  \bibinfo{author}{\bibfnamefont{G.}~\bibnamefont{Petekidis}},
  \bibinfo{journal}{Soft Matter} \textbf{\bibinfo{volume}{7}},
  \bibinfo{pages}{2456} (\bibinfo{year}{2011}).

\bibitem[{\citenamefont{Pham et~al.}(2006)\citenamefont{Pham, Petekidis,
  Vlassopoulos, Egelhaaf, Pusey, and Poon}}]{pham2006yielding}
\bibinfo{author}{\bibfnamefont{K.}~\bibnamefont{Pham}},
  \bibinfo{author}{\bibfnamefont{G.}~\bibnamefont{Petekidis}},
  \bibinfo{author}{\bibfnamefont{D.}~\bibnamefont{Vlassopoulos}},
  \bibinfo{author}{\bibfnamefont{S.}~\bibnamefont{Egelhaaf}},
  \bibinfo{author}{\bibfnamefont{P.}~\bibnamefont{Pusey}}, \bibnamefont{and}
  \bibinfo{author}{\bibfnamefont{W.}~\bibnamefont{Poon}}, \bibinfo{journal}{EPL
  (Europhysics Letters)} \textbf{\bibinfo{volume}{75}}, \bibinfo{pages}{624}
  (\bibinfo{year}{2006}).

\bibitem[{\citenamefont{Grenard et~al.}(2014)\citenamefont{Grenard, Divoux,
  Taberlet, and Manneville}}]{grenard2014timescales}
\bibinfo{author}{\bibfnamefont{V.}~\bibnamefont{Grenard}},
  \bibinfo{author}{\bibfnamefont{T.}~\bibnamefont{Divoux}},
  \bibinfo{author}{\bibfnamefont{N.}~\bibnamefont{Taberlet}}, \bibnamefont{and}
  \bibinfo{author}{\bibfnamefont{S.}~\bibnamefont{Manneville}},
  \bibinfo{journal}{Soft matter} \textbf{\bibinfo{volume}{10}},
  \bibinfo{pages}{1555} (\bibinfo{year}{2014}).

\bibitem[{\citenamefont{Leishangthem et~al.}(2017)\citenamefont{Leishangthem,
  Parmar, and Sastry}}]{Leishangthem2017Mar}
\bibinfo{author}{\bibfnamefont{P.}~\bibnamefont{Leishangthem}},
  \bibinfo{author}{\bibfnamefont{A.~D.~S.} \bibnamefont{Parmar}},
  \bibnamefont{and} \bibinfo{author}{\bibfnamefont{S.}~\bibnamefont{Sastry}},
  \bibinfo{journal}{Nat. Commun.} \textbf{\bibinfo{volume}{8}},
  \bibinfo{pages}{1} (\bibinfo{year}{2017}).

\bibitem[{\citenamefont{Karmakar et~al.}(2010)\citenamefont{Karmakar, Lerner,
  and Procaccia}}]{Karmakar2010Nov}
\bibinfo{author}{\bibfnamefont{S.}~\bibnamefont{Karmakar}},
  \bibinfo{author}{\bibfnamefont{E.}~\bibnamefont{Lerner}}, \bibnamefont{and}
  \bibinfo{author}{\bibfnamefont{I.}~\bibnamefont{Procaccia}},
  \bibinfo{journal}{Phys. Rev. E} \textbf{\bibinfo{volume}{82}},
  \bibinfo{pages}{055103} (\bibinfo{year}{2010}).

\bibitem[{\citenamefont{Maloney and
  Lema{\ifmmode\hat{\imath}\else\^{\i}\fi}tre}(2006)}]{Maloney2006Jul}
\bibinfo{author}{\bibfnamefont{C.~E.} \bibnamefont{Maloney}} \bibnamefont{and}
  \bibinfo{author}{\bibfnamefont{A.}~\bibnamefont{Lema{\ifmmode\hat{\imath}\else\^{\i}\fi}tre}},
  \bibinfo{journal}{Phys. Rev. E} \textbf{\bibinfo{volume}{74}},
  \bibinfo{pages}{016118} (\bibinfo{year}{2006}).

\bibitem[{\citenamefont{Keim and Arratia}(2013)}]{Keim2013}
\bibinfo{author}{\bibfnamefont{N.~C.} \bibnamefont{Keim}} \bibnamefont{and}
  \bibinfo{author}{\bibfnamefont{P.~E.} \bibnamefont{Arratia}},
  \bibinfo{journal}{Soft Matter} \textbf{\bibinfo{volume}{9}},
  \bibinfo{pages}{6222} (\bibinfo{year}{2013}).

\bibitem[{\citenamefont{Jaiswal et~al.}(2016)\citenamefont{Jaiswal, Procaccia,
  Rainone, and Singh}}]{Jaiswal2016Feb}
\bibinfo{author}{\bibfnamefont{P.~K.} \bibnamefont{Jaiswal}},
  \bibinfo{author}{\bibfnamefont{I.}~\bibnamefont{Procaccia}},
  \bibinfo{author}{\bibfnamefont{C.}~\bibnamefont{Rainone}}, \bibnamefont{and}
  \bibinfo{author}{\bibfnamefont{M.}~\bibnamefont{Singh}},
  \bibinfo{journal}{Phys. Rev. Lett.} \textbf{\bibinfo{volume}{116}},
  \bibinfo{pages}{085501} (\bibinfo{year}{2016}).

\bibitem[{\citenamefont{Regev et~al.}(2015)\citenamefont{Regev, Weber,
  Reichhardt, Dahmen, and Lookman}}]{Regev2015Nov}
\bibinfo{author}{\bibfnamefont{I.}~\bibnamefont{Regev}},
  \bibinfo{author}{\bibfnamefont{J.}~\bibnamefont{Weber}},
  \bibinfo{author}{\bibfnamefont{C.}~\bibnamefont{Reichhardt}},
  \bibinfo{author}{\bibfnamefont{K.~A.} \bibnamefont{Dahmen}},
  \bibnamefont{and} \bibinfo{author}{\bibfnamefont{T.}~\bibnamefont{Lookman}},
  \bibinfo{journal}{Nat. Commun.} \textbf{\bibinfo{volume}{6}},
  \bibinfo{pages}{1} (\bibinfo{year}{2015}).

\bibitem[{\citenamefont{Knowlton et~al.}(2014)\citenamefont{Knowlton, Pine, and
  Cipelletti}}]{Knowlton2014Aug}
\bibinfo{author}{\bibfnamefont{E.~D.} \bibnamefont{Knowlton}},
  \bibinfo{author}{\bibfnamefont{D.~J.} \bibnamefont{Pine}}, \bibnamefont{and}
  \bibinfo{author}{\bibfnamefont{L.}~\bibnamefont{Cipelletti}},
  \bibinfo{journal}{Soft Matter} \textbf{\bibinfo{volume}{10}},
  \bibinfo{pages}{6931} (\bibinfo{year}{2014}).

\bibitem[{\citenamefont{Shrivastav et~al.}(2016)\citenamefont{Shrivastav,
  Chaudhuri, and Horbach}}]{Shrivastav2016Oct}
\bibinfo{author}{\bibfnamefont{G.~P.} \bibnamefont{Shrivastav}},
  \bibinfo{author}{\bibfnamefont{P.}~\bibnamefont{Chaudhuri}},
  \bibnamefont{and} \bibinfo{author}{\bibfnamefont{J.}~\bibnamefont{Horbach}},
  \bibinfo{journal}{Phys. Rev. E} \textbf{\bibinfo{volume}{94}},
  \bibinfo{pages}{042605} (\bibinfo{year}{2016}).

\bibitem[{\citenamefont{Nicolas et~al.}(2018)\citenamefont{Nicolas, Ferrero,
  Martens, and Barrat}}]{nicolas2018deformation}
\bibinfo{author}{\bibfnamefont{A.}~\bibnamefont{Nicolas}},
  \bibinfo{author}{\bibfnamefont{E.~E.} \bibnamefont{Ferrero}},
  \bibinfo{author}{\bibfnamefont{K.}~\bibnamefont{Martens}}, \bibnamefont{and}
  \bibinfo{author}{\bibfnamefont{J.-L.} \bibnamefont{Barrat}},
  \bibinfo{journal}{Reviews of Modern Physics} \textbf{\bibinfo{volume}{90}},
  \bibinfo{pages}{045006} (\bibinfo{year}{2018}).

\bibitem[{\citenamefont{Sastry}(2020)}]{sastry2020models}
\bibinfo{author}{\bibfnamefont{S.}~\bibnamefont{Sastry}}
  (\bibinfo{year}{2020}), \eprint{arXiv: 2012.06726}.

\bibitem[{\citenamefont{Barlow et~al.}(2020)\citenamefont{Barlow, Cochran, and
  Fielding}}]{barlow2020ductile}
\bibinfo{author}{\bibfnamefont{H.~J.} \bibnamefont{Barlow}},
  \bibinfo{author}{\bibfnamefont{J.~O.} \bibnamefont{Cochran}},
  \bibnamefont{and} \bibinfo{author}{\bibfnamefont{S.~M.}
  \bibnamefont{Fielding}}, \bibinfo{journal}{Physical Review Letters}
  \textbf{\bibinfo{volume}{125}}, \bibinfo{pages}{168003}
  (\bibinfo{year}{2020}).

\bibitem[{\citenamefont{Donley et~al.}(2019)\citenamefont{Donley, de~Bruyn,
  McKinley, and Rogers}}]{donley2019time}
\bibinfo{author}{\bibfnamefont{G.~J.} \bibnamefont{Donley}},
  \bibinfo{author}{\bibfnamefont{J.~R.} \bibnamefont{de~Bruyn}},
  \bibinfo{author}{\bibfnamefont{G.~H.} \bibnamefont{McKinley}},
  \bibnamefont{and} \bibinfo{author}{\bibfnamefont{S.~A.}
  \bibnamefont{Rogers}}, \bibinfo{journal}{Journal of Non-Newtonian Fluid
  Mechanics} \textbf{\bibinfo{volume}{264}}, \bibinfo{pages}{117}
  (\bibinfo{year}{2019}).

\bibitem[{\citenamefont{Liu et~al.}(2021)\citenamefont{Liu, Dutta, Chaudhuri,
  and Martens}}]{liu2021elastoplastic}
\bibinfo{author}{\bibfnamefont{C.}~\bibnamefont{Liu}},
  \bibinfo{author}{\bibfnamefont{S.}~\bibnamefont{Dutta}},
  \bibinfo{author}{\bibfnamefont{P.}~\bibnamefont{Chaudhuri}},
  \bibnamefont{and} \bibinfo{author}{\bibfnamefont{K.}~\bibnamefont{Martens}},
  \bibinfo{journal}{Physical Review Letters} \textbf{\bibinfo{volume}{126}},
  \bibinfo{pages}{138005} (\bibinfo{year}{2021}).

\bibitem[{\citenamefont{Liu and Nagel}(1998)}]{Liu1998Nov}
\bibinfo{author}{\bibfnamefont{A.~J.} \bibnamefont{Liu}} \bibnamefont{and}
  \bibinfo{author}{\bibfnamefont{S.~R.} \bibnamefont{Nagel}},
  \bibinfo{journal}{Nature} \textbf{\bibinfo{volume}{396}}, \bibinfo{pages}{21}
  (\bibinfo{year}{1998}).

\bibitem[{\citenamefont{Fall et~al.}(2008)\citenamefont{Fall, Huang, Bertrand,
  Ovarlez, and Bonn}}]{Fall2008Jan}
\bibinfo{author}{\bibfnamefont{A.}~\bibnamefont{Fall}},
  \bibinfo{author}{\bibfnamefont{N.}~\bibnamefont{Huang}},
  \bibinfo{author}{\bibfnamefont{F.}~\bibnamefont{Bertrand}},
  \bibinfo{author}{\bibfnamefont{G.}~\bibnamefont{Ovarlez}}, \bibnamefont{and}
  \bibinfo{author}{\bibfnamefont{D.}~\bibnamefont{Bonn}},
  \bibinfo{journal}{Phys. Rev. Lett.} \textbf{\bibinfo{volume}{100}},
  \bibinfo{pages}{018301} (\bibinfo{year}{2008}).

\bibitem[{\citenamefont{Seto et~al.}(2019)\citenamefont{Seto, Singh,
  Chakraborty, Denn, and Morris}}]{Seto2019Aug}
\bibinfo{author}{\bibfnamefont{R.}~\bibnamefont{Seto}},
  \bibinfo{author}{\bibfnamefont{A.}~\bibnamefont{Singh}},
  \bibinfo{author}{\bibfnamefont{B.}~\bibnamefont{Chakraborty}},
  \bibinfo{author}{\bibfnamefont{M.~M.} \bibnamefont{Denn}}, \bibnamefont{and}
  \bibinfo{author}{\bibfnamefont{J.~F.} \bibnamefont{Morris}},
  \bibinfo{journal}{Granular Matter} \textbf{\bibinfo{volume}{21}},
  \bibinfo{pages}{82} (\bibinfo{year}{2019}).

\bibitem[{\citenamefont{V{\ifmmode\acute{a}\else\'{a}\fi}zquez-Quesada
  et~al.}(2016)\citenamefont{V{\ifmmode\acute{a}\else\'{a}\fi}zquez-Quesada,
  Tanner, and Ellero}}]{Vazquez-Quesada2016Aug}
\bibinfo{author}{\bibfnamefont{A.}~\bibnamefont{V{\ifmmode\acute{a}\else\'{a}\fi}zquez-Quesada}},
  \bibinfo{author}{\bibfnamefont{R.}~\bibnamefont{Tanner}}, \bibnamefont{and}
  \bibinfo{author}{\bibfnamefont{M.}~\bibnamefont{Ellero}},
  \bibinfo{journal}{Phys. Rev. Lett.} \textbf{\bibinfo{volume}{117}}
  (\bibinfo{year}{2016}).

\bibitem[{\citenamefont{Brown and Jaeger}(2014)}]{brown2014shear}
\bibinfo{author}{\bibfnamefont{E.}~\bibnamefont{Brown}} \bibnamefont{and}
  \bibinfo{author}{\bibfnamefont{H.~M.} \bibnamefont{Jaeger}},
  \bibinfo{journal}{Reports on Progress in Physics}
  \textbf{\bibinfo{volume}{77}}, \bibinfo{pages}{046602}
  (\bibinfo{year}{2014}).

\bibitem[{\citenamefont{Cheng et~al.}(2011)\citenamefont{Cheng, McCoy,
  Israelachvili, and Cohen}}]{Cheng2011Sep}
\bibinfo{author}{\bibfnamefont{X.}~\bibnamefont{Cheng}},
  \bibinfo{author}{\bibfnamefont{J.~H.} \bibnamefont{McCoy}},
  \bibinfo{author}{\bibfnamefont{J.~N.} \bibnamefont{Israelachvili}},
  \bibnamefont{and} \bibinfo{author}{\bibfnamefont{I.}~\bibnamefont{Cohen}},
  \bibinfo{journal}{Science} \textbf{\bibinfo{volume}{333}},
  \bibinfo{pages}{1276} (\bibinfo{year}{2011}).

\bibitem[{\citenamefont{Dhar et~al.}(2019)\citenamefont{Dhar, Chattopadhyay,
  and Majumdar}}]{Dhar2019Dec}
\bibinfo{author}{\bibfnamefont{S.}~\bibnamefont{Dhar}},
  \bibinfo{author}{\bibfnamefont{S.}~\bibnamefont{Chattopadhyay}},
  \bibnamefont{and} \bibinfo{author}{\bibfnamefont{S.}~\bibnamefont{Majumdar}},
  \bibinfo{journal}{J. Phys.: Condens. Matter} \textbf{\bibinfo{volume}{32}},
  \bibinfo{pages}{124002} (\bibinfo{year}{2019}).

\bibitem[{\citenamefont{Guy et~al.}(2015)\citenamefont{Guy, Hermes, and
  Poon}}]{guy2015towards}
\bibinfo{author}{\bibfnamefont{B.}~\bibnamefont{Guy}},
  \bibinfo{author}{\bibfnamefont{M.}~\bibnamefont{Hermes}}, \bibnamefont{and}
  \bibinfo{author}{\bibfnamefont{W.~C.} \bibnamefont{Poon}},
  \bibinfo{journal}{Physical review letters} \textbf{\bibinfo{volume}{115}},
  \bibinfo{pages}{088304} (\bibinfo{year}{2015}).

\bibitem[{\citenamefont{Koeze and Tighe}(2018)}]{Koeze2018Nov}
\bibinfo{author}{\bibfnamefont{D.~J.} \bibnamefont{Koeze}} \bibnamefont{and}
  \bibinfo{author}{\bibfnamefont{B.~P.} \bibnamefont{Tighe}},
  \bibinfo{journal}{Phys. Rev. Lett.} \textbf{\bibinfo{volume}{121}},
  \bibinfo{pages}{188002} (\bibinfo{year}{2018}).

\bibitem[{\citenamefont{Guy et~al.}(2018)\citenamefont{Guy, Richards, Hodgson,
  Blanco, and Poon}}]{guy2018constraint}
\bibinfo{author}{\bibfnamefont{B.}~\bibnamefont{Guy}},
  \bibinfo{author}{\bibfnamefont{J.}~\bibnamefont{Richards}},
  \bibinfo{author}{\bibfnamefont{D.}~\bibnamefont{Hodgson}},
  \bibinfo{author}{\bibfnamefont{E.}~\bibnamefont{Blanco}}, \bibnamefont{and}
  \bibinfo{author}{\bibfnamefont{W.}~\bibnamefont{Poon}},
  \bibinfo{journal}{Physical review letters} \textbf{\bibinfo{volume}{121}},
  \bibinfo{pages}{128001} (\bibinfo{year}{2018}).

\bibitem[{\citenamefont{Richards
  et~al.}(2020{\natexlab{a}})\citenamefont{Richards, Guy, Blanco, Hermes, Poy,
  and Poon}}]{Richards2020Mar}
\bibinfo{author}{\bibfnamefont{J.~A.} \bibnamefont{Richards}},
  \bibinfo{author}{\bibfnamefont{B.~M.} \bibnamefont{Guy}},
  \bibinfo{author}{\bibfnamefont{E.}~\bibnamefont{Blanco}},
  \bibinfo{author}{\bibfnamefont{M.}~\bibnamefont{Hermes}},
  \bibinfo{author}{\bibfnamefont{G.}~\bibnamefont{Poy}}, \bibnamefont{and}
  \bibinfo{author}{\bibfnamefont{W.~C.~K.} \bibnamefont{Poon}},
  \bibinfo{journal}{J. Rheol.} \textbf{\bibinfo{volume}{64}},
  \bibinfo{pages}{405} (\bibinfo{year}{2020}{\natexlab{a}}).

\bibitem[{\citenamefont{Richards
  et~al.}(2020{\natexlab{b}})\citenamefont{Richards, O{'}Neill, and
  Poon}}]{Richards2020Nov}
\bibinfo{author}{\bibfnamefont{J.~A.} \bibnamefont{Richards}},
  \bibinfo{author}{\bibfnamefont{R.~E.} \bibnamefont{O{'}Neill}},
  \bibnamefont{and} \bibinfo{author}{\bibfnamefont{W.~C.~K.}
  \bibnamefont{Poon}}, \bibinfo{journal}{Rheol. Acta} pp.
  \bibinfo{pages}{1--10} (\bibinfo{year}{2020}{\natexlab{b}}).

\bibitem[{\citenamefont{Singh et~al.}(2020{\natexlab{a}})\citenamefont{Singh,
  Ness, Seto, de~Pablo, and Jaeger}}]{Singh2020Jun}
\bibinfo{author}{\bibfnamefont{A.}~\bibnamefont{Singh}},
  \bibinfo{author}{\bibfnamefont{C.}~\bibnamefont{Ness}},
  \bibinfo{author}{\bibfnamefont{R.}~\bibnamefont{Seto}},
  \bibinfo{author}{\bibfnamefont{J.~J.} \bibnamefont{de~Pablo}},
  \bibnamefont{and} \bibinfo{author}{\bibfnamefont{H.~M.}
  \bibnamefont{Jaeger}}, \bibinfo{journal}{Phys. Rev. Lett.}
  \textbf{\bibinfo{volume}{124}}, \bibinfo{pages}{248005}
  (\bibinfo{year}{2020}{\natexlab{a}}).

\bibitem[{\citenamefont{Wyart and Cates}(2014)}]{wyart2014discontinuous}
\bibinfo{author}{\bibfnamefont{M.}~\bibnamefont{Wyart}} \bibnamefont{and}
  \bibinfo{author}{\bibfnamefont{M.}~\bibnamefont{Cates}},
  \bibinfo{journal}{Physical review letters} \textbf{\bibinfo{volume}{112}},
  \bibinfo{pages}{098302} (\bibinfo{year}{2014}).

\bibitem[{\citenamefont{Colombo and Del~Gado}(2014)}]{colombo2014stress}
\bibinfo{author}{\bibfnamefont{J.}~\bibnamefont{Colombo}} \bibnamefont{and}
  \bibinfo{author}{\bibfnamefont{E.}~\bibnamefont{Del~Gado}},
  \bibinfo{journal}{Journal of rheology} \textbf{\bibinfo{volume}{58}},
  \bibinfo{pages}{1089} (\bibinfo{year}{2014}).

\bibitem[{\citenamefont{van Doorn et~al.}(2018)\citenamefont{van Doorn,
  Verweij, Sprakel, and van~der Gucht}}]{vanDoorn2018May}
\bibinfo{author}{\bibfnamefont{J.~M.} \bibnamefont{van Doorn}},
  \bibinfo{author}{\bibfnamefont{J.~E.} \bibnamefont{Verweij}},
  \bibinfo{author}{\bibfnamefont{J.}~\bibnamefont{Sprakel}}, \bibnamefont{and}
  \bibinfo{author}{\bibfnamefont{J.}~\bibnamefont{van~der Gucht}},
  \bibinfo{journal}{Phys. Rev. Lett.} \textbf{\bibinfo{volume}{120}},
  \bibinfo{pages}{208005} (\bibinfo{year}{2018}).

\bibitem[{\citenamefont{B{\ifmmode\acute{e}\else\'{e}\fi}cu
  et~al.}(2006)\citenamefont{B{\ifmmode\acute{e}\else\'{e}\fi}cu, Manneville,
  and Colin}}]{Becu2006Apr}
\bibinfo{author}{\bibfnamefont{L.}~\bibnamefont{B{\ifmmode\acute{e}\else\'{e}\fi}cu}},
  \bibinfo{author}{\bibfnamefont{S.}~\bibnamefont{Manneville}},
  \bibnamefont{and} \bibinfo{author}{\bibfnamefont{A.}~\bibnamefont{Colin}},
  \bibinfo{journal}{Phys. Rev. Lett.} \textbf{\bibinfo{volume}{96}},
  \bibinfo{pages}{138302} (\bibinfo{year}{2006}).

\bibitem[{\citenamefont{Liberto et~al.}(2020)\citenamefont{Liberto, Le~Merrer,
  Manneville, and Barentin}}]{Liberto2020Oct}
\bibinfo{author}{\bibfnamefont{T.}~\bibnamefont{Liberto}},
  \bibinfo{author}{\bibfnamefont{M.}~\bibnamefont{Le~Merrer}},
  \bibinfo{author}{\bibfnamefont{S.}~\bibnamefont{Manneville}},
  \bibnamefont{and} \bibinfo{author}{\bibfnamefont{C.}~\bibnamefont{Barentin}},
  \bibinfo{journal}{Soft Matter} \textbf{\bibinfo{volume}{16}},
  \bibinfo{pages}{9217} (\bibinfo{year}{2020}).

\bibitem[{\citenamefont{Irani et~al.}(2014)\citenamefont{Irani, Chaudhuri, and
  Heussinger}}]{Irani2014May}
\bibinfo{author}{\bibfnamefont{E.}~\bibnamefont{Irani}},
  \bibinfo{author}{\bibfnamefont{P.}~\bibnamefont{Chaudhuri}},
  \bibnamefont{and}
  \bibinfo{author}{\bibfnamefont{C.}~\bibnamefont{Heussinger}},
  \bibinfo{journal}{Phys. Rev. Lett.} \textbf{\bibinfo{volume}{112}},
  \bibinfo{pages}{188303} (\bibinfo{year}{2014}).

\bibitem[{\citenamefont{Irani et~al.}(2016)\citenamefont{Irani, Chaudhuri, and
  Heussinger}}]{Irani2016Nov}
\bibinfo{author}{\bibfnamefont{E.}~\bibnamefont{Irani}},
  \bibinfo{author}{\bibfnamefont{P.}~\bibnamefont{Chaudhuri}},
  \bibnamefont{and}
  \bibinfo{author}{\bibfnamefont{C.}~\bibnamefont{Heussinger}},
  \bibinfo{journal}{Phys. Rev. E} \textbf{\bibinfo{volume}{94}},
  \bibinfo{pages}{052608} (\bibinfo{year}{2016}).

\bibitem[{\citenamefont{Chaudhuri et~al.}(2012)\citenamefont{Chaudhuri,
  Berthier, and Bocquet}}]{Chaudhuri2012Feb}
\bibinfo{author}{\bibfnamefont{P.}~\bibnamefont{Chaudhuri}},
  \bibinfo{author}{\bibfnamefont{L.}~\bibnamefont{Berthier}}, \bibnamefont{and}
  \bibinfo{author}{\bibfnamefont{L.}~\bibnamefont{Bocquet}},
  \bibinfo{journal}{Phys. Rev. E} \textbf{\bibinfo{volume}{85}},
  \bibinfo{pages}{021503} (\bibinfo{year}{2012}).

\bibitem[{\citenamefont{Vasisht et~al.}(2020)\citenamefont{Vasisht, Roberts,
  and Del~Gado}}]{vasisht2020emergence}
\bibinfo{author}{\bibfnamefont{V.~V.} \bibnamefont{Vasisht}},
  \bibinfo{author}{\bibfnamefont{G.}~\bibnamefont{Roberts}}, \bibnamefont{and}
  \bibinfo{author}{\bibfnamefont{E.}~\bibnamefont{Del~Gado}},
  \bibinfo{journal}{Physical Review E} \textbf{\bibinfo{volume}{102}},
  \bibinfo{pages}{010604} (\bibinfo{year}{2020}).

\bibitem[{\citenamefont{Singh et~al.}(2020{\natexlab{b}})\citenamefont{Singh,
  Ozawa, and Berthier}}]{singh2020brittle}
\bibinfo{author}{\bibfnamefont{M.}~\bibnamefont{Singh}},
  \bibinfo{author}{\bibfnamefont{M.}~\bibnamefont{Ozawa}}, \bibnamefont{and}
  \bibinfo{author}{\bibfnamefont{L.}~\bibnamefont{Berthier}},
  \bibinfo{journal}{Physical Review Materials} \textbf{\bibinfo{volume}{4}},
  \bibinfo{pages}{025603} (\bibinfo{year}{2020}{\natexlab{b}}).

\bibitem[{\citenamefont{Trappe et~al.}(2001)\citenamefont{Trappe, Prasad,
  Cipelletti, Segre, and Weitz}}]{Trappe2001Jun}
\bibinfo{author}{\bibfnamefont{V.}~\bibnamefont{Trappe}},
  \bibinfo{author}{\bibfnamefont{V.}~\bibnamefont{Prasad}},
  \bibinfo{author}{\bibfnamefont{L.}~\bibnamefont{Cipelletti}},
  \bibinfo{author}{\bibfnamefont{P.~N.} \bibnamefont{Segre}}, \bibnamefont{and}
  \bibinfo{author}{\bibfnamefont{D.~A.} \bibnamefont{Weitz}},
  \bibinfo{journal}{Nature} \textbf{\bibinfo{volume}{411}},
  \bibinfo{pages}{772} (\bibinfo{year}{2001}).

\bibitem[{\citenamefont{Maxwell}(1864)}]{Maxwell1864Apr}
\bibinfo{author}{\bibfnamefont{J.~C.} \bibnamefont{Maxwell}},
  \bibinfo{journal}{London, Edinburgh, and Dublin Philosophical Magazine and
  Journal of Science} \textbf{\bibinfo{volume}{27}}, \bibinfo{pages}{294}
  (\bibinfo{year}{1864}).

\bibitem[{\citenamefont{Peters et~al.}(2016)\citenamefont{Peters, Majumdar, and
  Jaeger}}]{Peters2016Apr}
\bibinfo{author}{\bibfnamefont{I.~R.} \bibnamefont{Peters}},
  \bibinfo{author}{\bibfnamefont{S.}~\bibnamefont{Majumdar}}, \bibnamefont{and}
  \bibinfo{author}{\bibfnamefont{H.~M.} \bibnamefont{Jaeger}},
  \bibinfo{journal}{Nature} \textbf{\bibinfo{volume}{532}},
  \bibinfo{pages}{214} (\bibinfo{year}{2016}).

\bibitem[{\citenamefont{Storm et~al.}(2005)\citenamefont{Storm, Pastore,
  MacKintosh, Lubensky, and Janmey}}]{Storm2005May}
\bibinfo{author}{\bibfnamefont{C.}~\bibnamefont{Storm}},
  \bibinfo{author}{\bibfnamefont{J.~J.} \bibnamefont{Pastore}},
  \bibinfo{author}{\bibfnamefont{F.~C.} \bibnamefont{MacKintosh}},
  \bibinfo{author}{\bibfnamefont{T.~C.} \bibnamefont{Lubensky}},
  \bibnamefont{and} \bibinfo{author}{\bibfnamefont{P.~A.}
  \bibnamefont{Janmey}}, \bibinfo{journal}{Nature}
  \textbf{\bibinfo{volume}{435}}, \bibinfo{pages}{191} (\bibinfo{year}{2005}).

\end{thebibliography}

\end{document}